\pdfoutput=1
\documentclass[preprint]{aastex}
\usepackage{lscape} 
\usepackage{amsmath, amsthm,amssymb}
\usepackage{multirow}
\usepackage{url}

\newcommand{\FOP}{ \textbf{FOP}-test }

\urldef\mboss\url{http://www.eso.org/~ohainaut/MBOSS}

\title{The Hubble Wide Field Camera 3 Test of Surfaces in the Outer Solar System: The Compositional Classes of the Kuiper Belt}

\author{Wesley C. Fraser {$^1$}}
\altaffiltext{1}{Division of Geological and Planetary Sciences, California Institute of Technology 1200 E. California Blvd. Pasadena, CA USA 91125}
\email{fraserw@gps.caltech.edu}
\author{Michael E. Brown {$^1$}}
\date{} 

\slugcomment{Submitted to ApJ, 08/18/11. Revised 11/14/11}
\received{August 18, 2011}
\revised{November 30, 2011}

\begin{abstract}
We present the first results of the Hubble Wide Field Camera 3 Test of Surfaces in the Outer Solar System (H/WTSOSS). The purpose of this survey was to measure the surface properties of a large number of Kuiper belt objects and attempt to infer compositional and dynamical correlations. We find that the Centaurs and the low-perihelion scattered disk and resonant objects exhibit virtually identical bifurcated optical colour distributions and make up two well defined groups of object. Both groups have highly correlated optical and NIR colours which are well described by a pair of two component mixture models that have different red components, but share a common neutral component. The small, $H_{606}\gtrsim5.6$ high-perihelion excited objects are entirely consistent with being drawn from the two branches of the mixing model suggesting that the colour bifurcation of the Centaurs is apparent in all small excited objects. On the other hand, objects larger than $H_{606}\sim5.6$ are not consistent with the mixing model, suggesting some evolutionary process avoided by the smaller objects. The existence of a bifurcation amongst all excited populations argues that the two separate classes of object existed in the primordial disk before the excited Kuiper belt was populated. The cold classical objects exhibit a different type of surface which has colours that are consistent with being drawn from the red branch of the mixing model, but with much higher albedos.
\end{abstract}

\begin{document}

\maketitle

\section{Introduction \label{sec:Intro}}
Kuiper belt objects (KBOs), planetesimals near and beyond the orbit of Neptune, are debris left over from the planet formation process that occurred in the early Solar system. Because of the relatively large heliocentric distances, KBOs likely present some of the most pristine material available for astronomical study, and as such, the Kuiper belt is often regarded as a sort of archaeological window into the early history of the outer Solar system. 

Many KBOs are trapped in mean-motion resonances with Neptune, or are on highly inclined and eccentric orbits \citep{Trujillo2001}. This has been taken as strong evidence that the majority of KBO orbits are not primordial, but rather these objects likely formed closer to the Sun, and have since been scattered onto their current orbits. Resonance sweeping during the smooth migration of Neptune was originally put forth as a possible mechanism responsible for populating the mean-motion resonances \citep{Malhotra1993}. This process however, typically produces a Neptune with too low an eccentricity \citep[see][for a review]{Morbidelli2008}. An alternative, and the currently favoured process, is scattering and chaotic capture by an eccentric Neptune \citep[see for example][]{Levison2008}.

It seems that the  Kuiper belt's current members have experienced a diverse range of dynamical histories, and were delivered  from broad range of heliocentric distances. It has been the hope that a compositional study of KBOs would reveal a wealth of information, not just about their chemical make-up, but also about their past dynamical path-ways and relative formation locations within the primordial disk.

Spectroscopic studies of KBOs have revealed that these objects fall into three general classes. Members of the Haumea collisional family, of which the KBO Haumea (136108) is the largest,  all exhibit spectra with the deepest water-ice absorption features seen in the Kuiper belt consistent with surfaces of nearly pure water-ice \citep{Brown2007}. Non-Haumea family members make up the remaining two classes. The largest KBOs are found to be volatile rich, bearing ices such as methane and nitrogen \citep{Cruikshank1976,Schaller2007b}. The spectra of smaller objects are found to be linear in the wavelength range $\lambda \sim0.5 \mbox{ - }0.9 \mbox{ $\mu$m}$ \citep{Hainaut2002,Fornasier2009} with varying optical colours. This slope flattens towards more neutral colours at $\sim 0.9-1 \mbox{ $\mu$m}$. The small KBOs are all found to be absent of any absorptions due to the volatile ices exhibited by the larger objects and might only show some amount of water-ice absorption \citep{Barkume2008,Guilbert2009}. This is in agreement with the volatile retention model presented by \citet{Schaller2007a} which demonstrates that only the largest objects have sufficient gravity to retain their volatiles against sublimation over the age of the Solar system.

While the lack of volatile ices on the surfaces of the smallest objects is understood, the colours of these objects remain enigmatic. Optical surveys have revealed two peculiar correlations between optical colour and orbital elements. The Centaurs, objects which reside on orbits between Saturn and Neptune, exhibit a bimodality in their optical colours, and fall into two distinctive groups, the blue group has optical colours $1\lesssim \mbox{(B-R)}\lesssim1.4$, and the red group has $1.6\lesssim \mbox{(B-R)}\lesssim2$ with no known objects in between \citep{Tegler2003,Tegler2008}. 

The classical objects, which reside on orbits between the 2:3 and 1:2 exterior mean-motion resonances with Neptune ($40\lesssim a \lesssim48$) and have low to moderate eccentricities exhibit nearly unanimously red colours at low inclinations. At higher inclinations on the other hand, the classical objects exhibit a broad range of colours with much bluer objects than those found at low inclinations \citep{Doressoundiram2002,Peixinho2004,Peixinho2008}.

It might be expected that a two component model of water-ice and some red organic material would explain the surfaces of the small KBOs. \citet{Luu1996,Jewitt2001} presented a model in which the surfaces of KBOs were governed by two competing processes. Cosmic-ray bombardment over time would lead to a dark red irradiation mantel  rich in complex organic material, while impact gardening would dredge up fresh blue, non-irradiated ice. If successful, this simple model would imply  that the optical colours, amount of water-ice absorption, and geometric albedos of KBOs would be correlated. It has been shown that this is not the case \citep{Barkume2008}. In addition, it has been shown that both components in binary KBO systems exhibit virtually identical optical colours \citep{Benecchi2009} incompatible with such a model. No model to date has convincingly accounted for the surface properties exhibited by the small KBOs.

We have performed a large spectro-photometric survey with the Hubble Space Telescope (Cycle 17, HST-GO-Program 11644). The survey, designated the Hubble/WFC3 Test of Surfaces in the Outer Solar System (H/WTSOSS) was designed to take advantage of the deep optical and NIR imaging capabilities introduced with the Wide Field Camera 3 (WFC3) to obtain high-quality measurements of both the optical colour and water-ice absorption on a large sample of KBOs too faint to be observed with similar quality in the NIR from even the largest ground-based offerings. We present the primary results of this project here.

In section~\ref{sec:observations} we present our sample selection, observational techniques, and data reduction methods. In section~\ref{sec:results} we present the results of our photometry. We find that the targets in our survey fall generally into two classes based on their surfaces and orbits. In section~\ref{sec:discussion}  we discuss what insights can be gained from our observations about the formation of the Kuiper belt. Finally, we present our conclusions in  section~\ref{sec:conclusions}.

\section{Observations and Data Reductions \label{sec:observations}}
In this section we present our target sample and observation techniques. Additionally, we present our data reductions and the resultant photometry acquired by the H/WTSOSS survey. The results of this survey along with previously published HST photometry of KBOs will be available online in database form at \url{www.fraserkbos.com}. 

\subsection{Target Selection and Classification \label{sec:targets}}
The primary goal of this survey was to gain optical and NIR spectroscopic information on KBOs too faint to be observed from the ground. Past NIR spectrocscopic surveys have demonstrated an approximate ground-based magnitude limit of V$\sim21$ \citep{Barkume2008,Guilbert2009,Barucci2011}. As such, with only a few exceptions, we restrict ourselves to objects with brightnesses below this level. Signal to noise estimates suggested that good quality photometry could be acquired for objects as faint as V$\sim24$, and we used this to bound our target list.

\begin{figure}[h]
   \centering
   \plotone{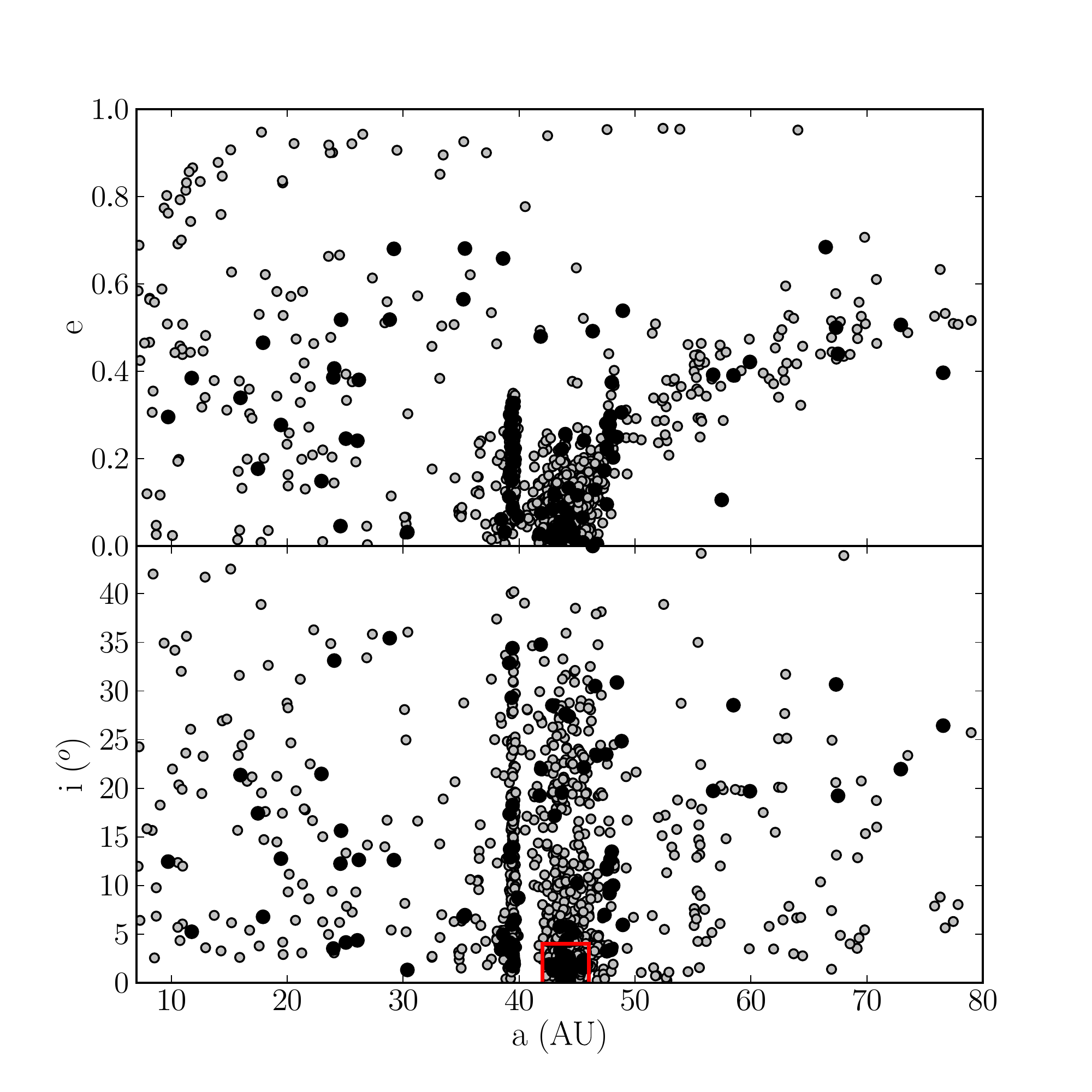} 
   \figcaption{Eccentricity and inclination versus semi-major axis of objects in the MPC \citep{MPCORB} (grey points) and the targets in this survey (black points). The red box shows the element space of the objects we designate as cold classical KBOs. \label{fig:aei}}
\end{figure}

Effort was made to observe as diverse a range of KBOs as possible. Targets were selected from the Minor Planet Center (MPC) \citep{MPCORB} based on their orbital quality to maximize the chances that the object would fall in the WFC3 field of view during cycle 17. The result was a selection of 120 targets spanning the full range of orbital element space occupied by the Kuiper belt. The observed targets are presented in Table~\ref{tab:BIGTABLE}. As well, we present the orbital elements of our targets in Figure~\ref{fig:aei}.

For completeness and historical reasons, we include orbital classifications for the observed objects. As shall be seen however, the adopted orbital classification scheme has little bearing on the results of this work. We adopt an orbital classification scheme inspired by the system of \citet{Lykawka2007}.  We define Centaurs as those objects having semi-major axes, $a<30$ AU or perihelia, $q<25$ AU. As well, we define resonant objects as those objects in the 1:2, 2:3, 3:5, or 4:7 mean-motion resonances with Neptune. 

It is well accepted that the non-resonant KBOs with semi-major axes bounded roughly by the 2:3 and 1:2 mean-motion resonances divide in inclination, $i$, into two populations \citep{Brown2001}. Historically, these two populations, the hot and cold classical objects, are defined as those objects with inclinations above and below $i=5^o$ respectively \citep{Elliot2005}. The cold classical objects (CCOs) which all exhibit similar red surfaces, are on orbits with moderate eccentricities $e\lesssim 0.15$, while the hot classical objects (HCOs) exhibit a more diverse range of colours and eccentricities \citep[see for example][]{Peixinho2008}. From a colour standpoint, the inclination dividing these two populations is ambiguous. Dynamically however, the inclination division is clear. Analysis of Figure~\ref{fig:aei} reveals a tight clump of objects with $i<4^o$, and $42\lesssim a \lesssim 46$. For $i<4^o$, very few known non-resonant objects fall outside of this range in semi-major axis; none of these were observed in our survey. Thus, we adopt a definition for the CCOs, as those objects with $i<4$ and $42<a<46$. We define HCOs as non-resonant objects with $a$ bounded by the 1:2 and 2:3 mean-motion resonances, $i>4^o$, and $q>37$ AU.

Those remaining objects, not defined as Centaurs, resonant, CCOs, or HCOs are classified as scattered objects. It is worth noting that in some other works the scattered population is divided into groups: those objects on orbits which allow them to have interactions with neptune, and those which are now well isolated from Neptune \citep[see][]{Gladman2008}. Only 2 of our survey targets, which we classify as scattered, are not on interaction orbits. Therefore,  we avoid this additional sub-classification as an unnecessary complication for the purposes of this manuscript. Modifications to our classifications or the adoption of dynamically motivated and more complicated schemes such as that presented by \citet{Gladman2008} do not affect our conclusions.

\subsection{Observations and Data Reductions \label{sec:observing}}
The primary scientific purpose of H/WTSOSS was to determine optical slopes and amount of water-ice absorption on faint KBOs. The excellent photometric stability of HST along with the imaging capabilities  and filter selection of WFC3 made this new camera the ideal instrument for the job. 

Each target was observed in exactly the same way. During the observations the telescope was slewed at the apparent rate of motion of the target. Each target was visited only once and monitored for the entire visibility window of a single orbit. This ensured that no significant rotational variability would affect our photometry between different filters. Pairs of images were taken in the F606w, F814w, F139m, and F153m bandpasses with exposure times of 129 s and 453 s in each of the optical and NIR filters respectively. Dithers of 3'' were included between each image of a pair to ensure that the photometry was not affected by cosmic rays or bad regions of a detector. The detectors were windowed to minimize overheads and to ensure the data could be transferred from the telescope during Earth occultation. This resulted in small square fields of view of 20.5" and 131" on a side for the optical and NIR images. As a result, 8 objects fell outside the optical windows and were only seen in the NIR images. In 6 cases, the objects were not seen in the optical or NIR images entirely. In addition,  6 objects fell near enough to bright background sources to prevent reliable photometry in at least a few images. These images were excluded from our analysis (see Table~\ref{tab:BIGTABLE}).

Images were first processed through the standard WFC3 data processing pipeline, CalWF3 \citep{Rajan2010}. As of publication, all data presented here were processed using version 2.3 of this pipeline. Images were then visually inspected and bad pixels, and cosmic ray strikes not identified during the initial processing were flagged.

Analysis of the {\it Tiny Tim} version 7.1 \citep{Krist1993} simulated point-spread functions (PSFs) revealed that the simulated PSFs did an adequate job of reproducing the outermost features of the observed PSFs (first airy ring, diffraction spikes, etc.) but failed to reproduce the image cores. As the majority of source flux is contained within the PSF core, photometry derived from matching {\it Tiny Tim} PSFs to the observations was deemed unreliable. 

Photometry was derived using standard aperture photometry techniques. Aperture radii were chosen as a function of source brightness and wavelength to minimize measurement scatter. To minimize the effects of bad pixels and cosmic rays, {\it Tiny Tim} PSFs were used to interpolate over any bad pixels identified in the CalWF3 processing. Typically, only 1 pixel per aperture was flagged as bad. In addition, the simulated PSFs were used to derive infinite aperture corrections for each object. As the aperture radii were kept larger than the radius of the first airy ring,  any errors introduced from the simulated PSF cores were negligible.  

In a few cases, partially resolved binary sources were identified. In those cases, an aperture large enough to encompass both sources was used, and the combined photometry of both sources is reported.  The adopted photometric system is the STMAG system. The STMAG system defines an equivalent flux density for a source with spectrum that is flat as a function of wavelength that would produce the observed count rate. The magnitude in this system is just $STMAG = -2.5 \log f_\lambda -Z$ where $Z$ is the filter zero point and $f_\lambda$ is the flux density expressed in erg cm$^{-2}$ sec$^{-1}$ \AA$^{-1}$ The resultant photometry are presented in Table~\ref{tab:BIGTABLE} along with object colours in Table~\ref{tab:BIGTABLE2}.

Object 2000 EE173 (60608) fell within the IR field of view, but outside the UVIS field of view during observations. Optical data from \citet{Benecchi2011} in the F675w and F814w filters was available for this from which a reliable estimate of the F606w, and (F606w-F814w) colour could be made. Conversion was done using an analog Solar star and the {\it synphot} routine. We note that by estimating the F606w flux of 2000 EE173 in this way may introduce errors due to rotational variability. \citet{Duffard2009} has shown that the majority of non-Centaur KBOs have peak-to-peak lightcurve amplitudes of $\lesssim0.1$ mags. As such, the photometric errors reported by \citet{Benecchi2011} were set to 0.1 mags to roughly account for any errors in estimating the F606w flux.  This result is presented along with our other photometry in  Table\ref{tab:BIGTABLE2}. No IR observations in the literature could be used to accurately determine F139m and F153m photometry of those objects which fell outside the IR field of view.

\subsection{Photometric Uncertainties}
\begin{figure}[h] 
   \centering
   \plotone{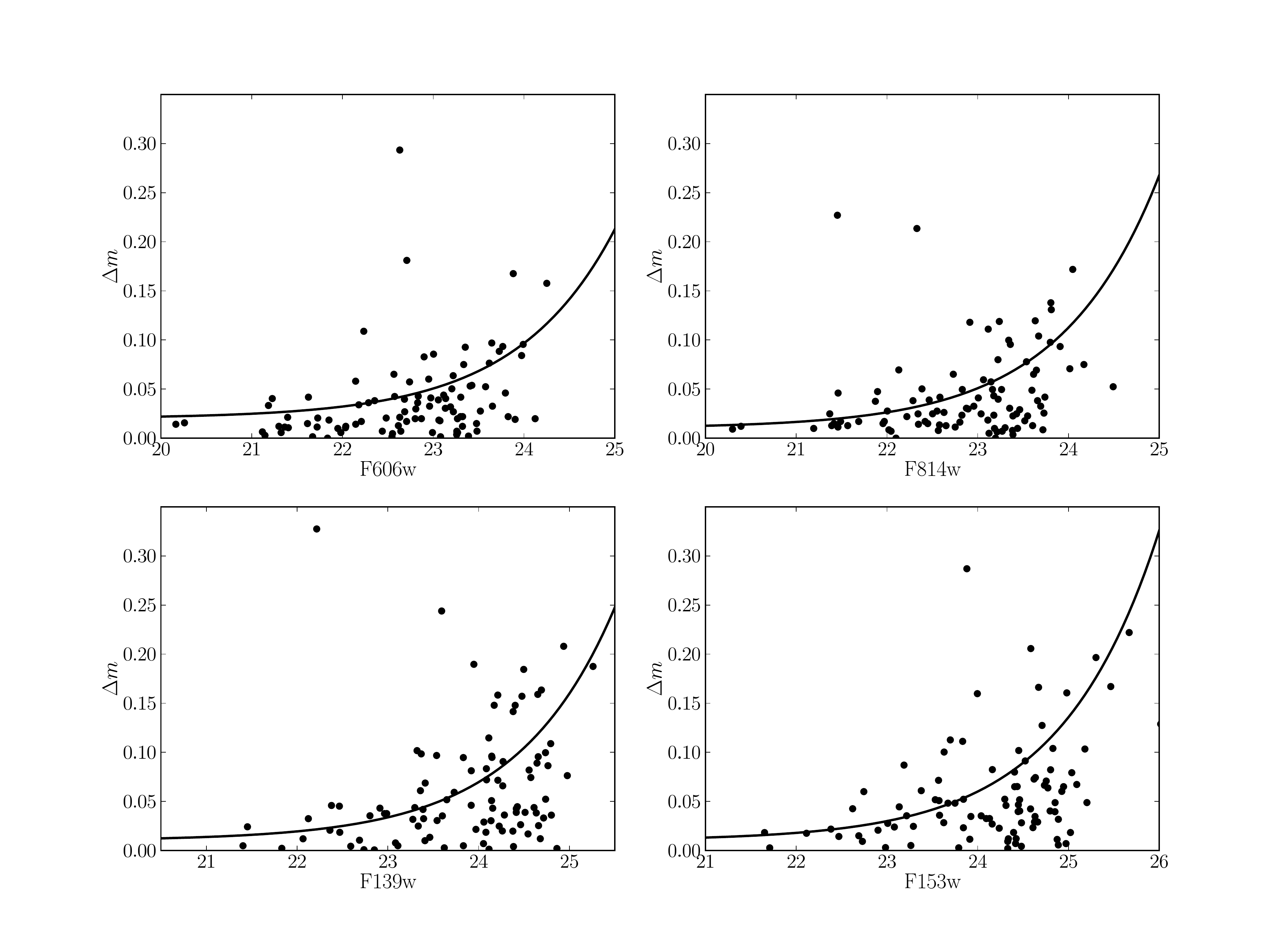} 
   \figcaption{Scatter in the photometry from the image dither pairs versus the average in the two measurements for all four filters. The curves represent the fit of Equation~\ref{eq:photoUncertainty} to the presented data. Parameters $(C,\gamma,Z)$ for these curves are $(0.02,0.6,26.234)$, $(0.01,0.6,25.918)$, $(0.01,0.6,26.507)$, and $(0.01,0.6,26.697)$ for the F606w, F814w, F139m, and F153m filters respectively. \label{fig:photoScatter}}
\end{figure}

In the case of photometry in either the flux or background limited regimes, the uncertainty in the observed magnitude, $m$ of a source can be well approximated by 

\begin{equation}
\Delta m \sim \gamma 10^{\frac{m-Z}{2.5}}
\end{equation}

\noindent 
where $Z$ is the photometric zeropoint of the passband in question. Theoretically, the proportionality constant $\gamma$ is a function of the telescope and detector parameters such as gain, read-noise, and data reductions parameters such as aperture size. In practice, this parameter is a factor of a few larger than the theoretical expectation \citep[see for instance][]{Newberry1991,Fraser2008}. In addition, theory underestimates the photometric scatter in repeated measurements of bright objects (see Figure~\ref{fig:photoScatter}). As a result, we consider an additional constant, $C$ and represent our photometric uncertainty by 

\begin{equation}
\Delta m = C + \gamma 10^{\frac{m-Z}{2.5}}.
\label{eq:photoUncertainty}
\end{equation}

\noindent
This equation is fit to the observed scatter from the dithered images of each source and each passband. The results of this procedure along with the best-fit parameters are presented in Figure~\ref{fig:photoScatter}. As can be seen, Equation~\ref{eq:photoUncertainty} provides an adequate description of the measurement scatter in the photometry. As well, it can be seen that approximately 1\% of measurements deviate significantly away from the curves given by Equation~\ref{eq:photoUncertainty} primarily as a result of cosmic ray strikes affecting one of the measurements of a pair. Thus, if the difference in pairs of observed magnitudes for a given filter was larger than 1.4 times the direct sum of errors of the pair of measurements, then it was assumed that the brighter of the two measurements was affected by a nearby cosmic ray strike and was subsequently rejected. 

The photometry presented here are average values from the pairs in each filter, or the single value when a measurement was deemed bad, along with photometric uncertainties determined from the best-fit values of Equation~\ref{eq:photoUncertainty} added in quadrature; the quoted photometric uncertainties represent 1-$\sigma$ errors. 

\section{Results \label{sec:results}}
In this section we present the three colour photometric results of  the H/WTSOSS observations along with an interpretation of the observations in terms of a two component mixing model. The main result is the identification of three types of KBO surface.

\subsection{The Low Perihelia Objects \label{sec:low_q}}

\begin{figure}[h] 
   \centering
      \epsscale{0.8}
      \plotone{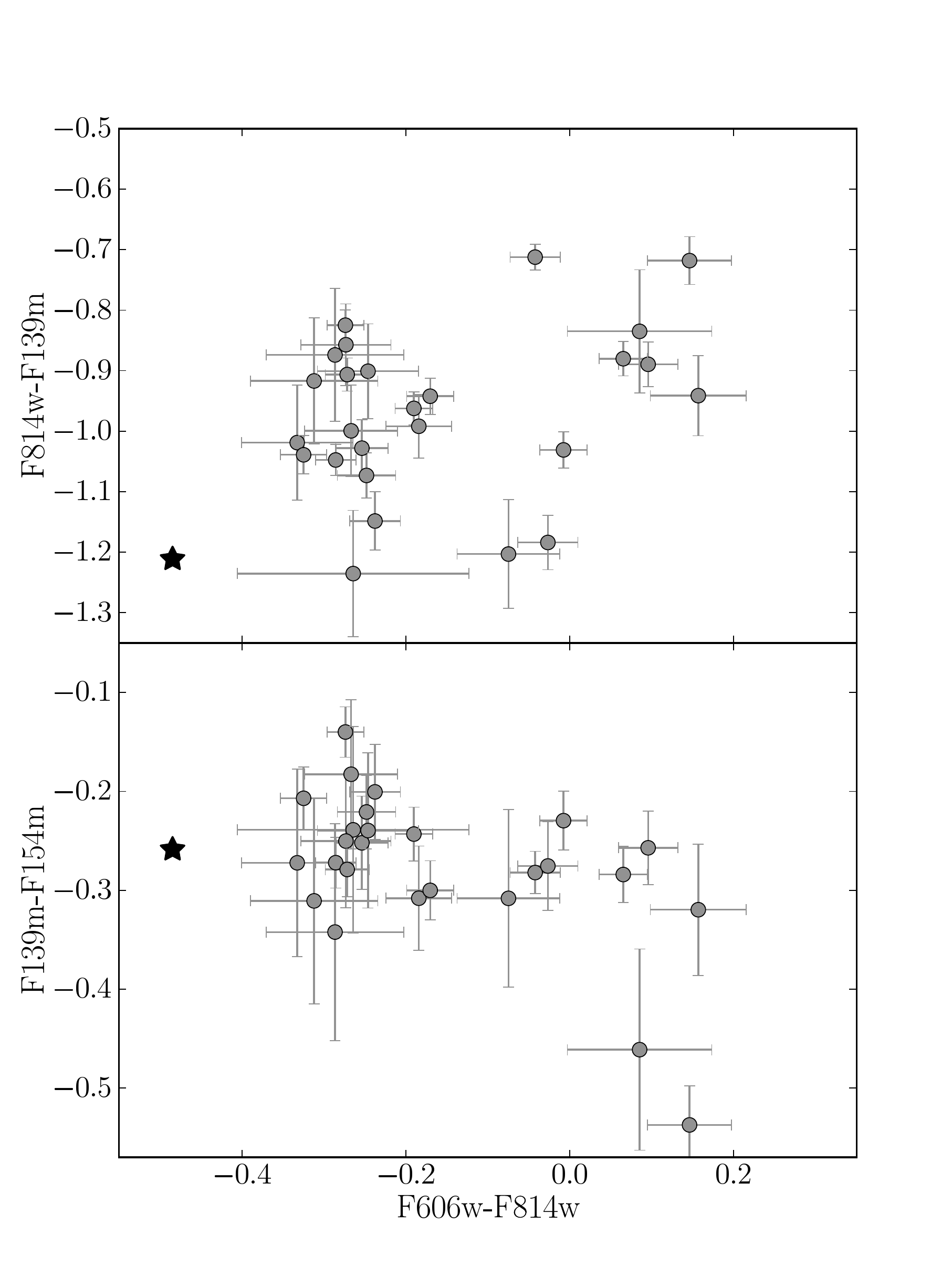} 
   \figcaption{(F814w-F139m) and (F139m-F154m) vs. (F606w-F814w) colours of the Centaurs. Solar colours, (F606w-F814w)$_{\bigodot}=-0.48$, (F814w-F139m)$_{\bigodot}=-1.21$, and (F139m-F154m)$_{\bigodot}=-0.26$ are shown by the star. Only those objects with good measurements in all four filters are shown.  \label{fig:CC_centaurs_noBoxes}}
\end{figure}

\begin{figure}[h] 
   \centering
      \epsscale{0.8}
      \plotone{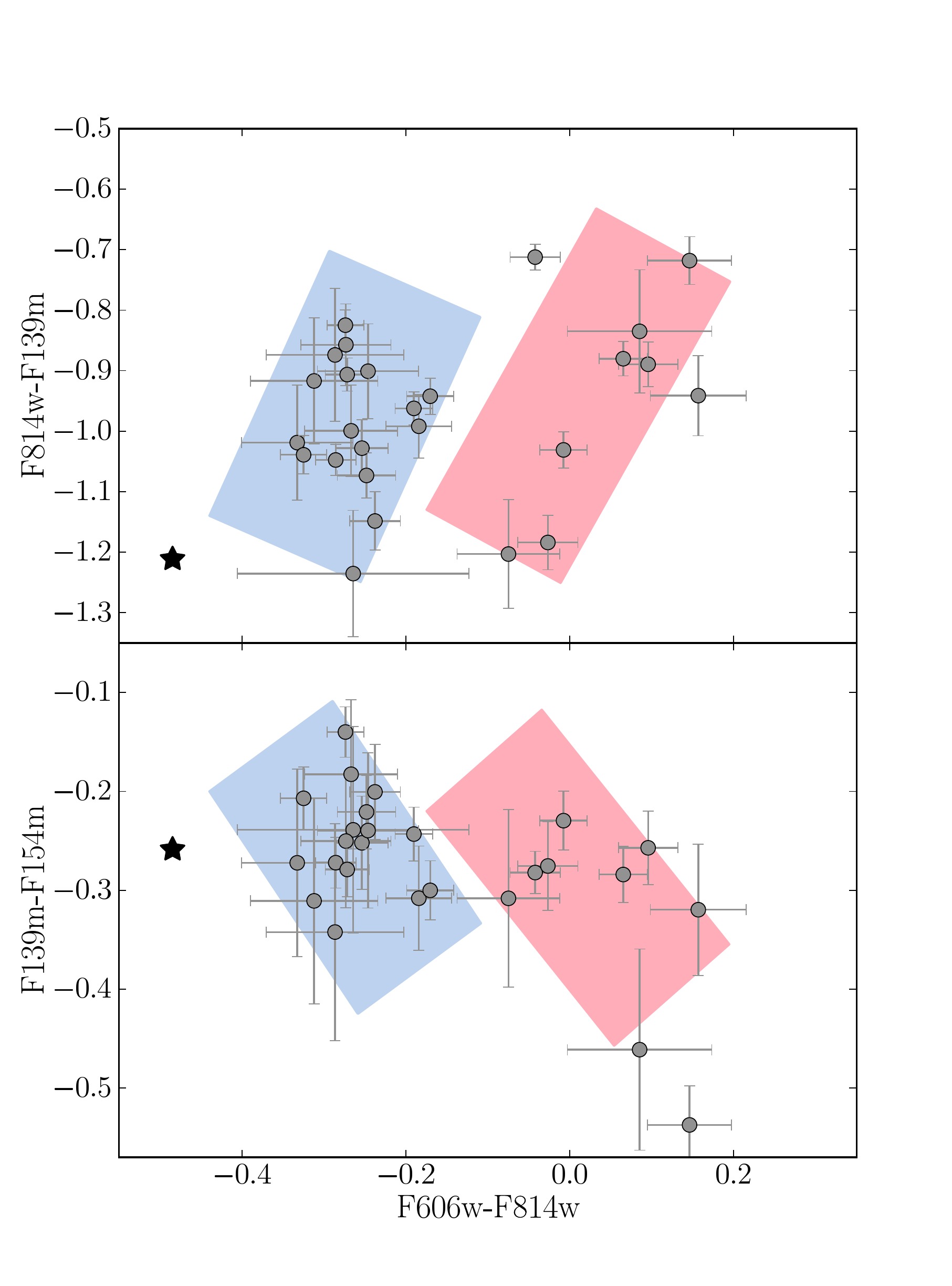} 
   \figcaption{(F814w-F139m) and (F139m-F154m) vs. (F606w-F814w) colours of the Centaurs. Solar colours, (F606w-F814w)$_{\bigodot}=-0.48$, (F814w-F139m)$_{\bigodot}=-1.21$, and (F139m-F154m)$_{\bigodot}=-0.26$ are shown by the star. Only those objects with good measurements in all four filters are shown. The extent of the red and blue shaded regions are selected to span the approximate extent of the red and blue classes of low-$q$ object (see Section~\ref{sec:low_q}). These regions and the axis limits are the same for Figures~\ref{fig:CC_centaurs}, \ref{fig:CC_qlt35}, \ref{fig:CC_qgt35}, \ref{fig:CC_qgt35_wModel}, and \ref{fig:CC_cc} to ease comparison between them. \label{fig:CC_centaurs}}
\end{figure}

In Figures~\ref{fig:CC_centaurs_noBoxes} and \ref{fig:CC_centaurs} we present the observed (F606w-F814w), (F814w-F193m), and (F139m-F154m)  colours of the Centaurs. The well known bifurcation of the optical colours of the Centaurs can be seen with the two colour groups bifurcating about an optical colour of (F606w-F814w)$\sim-0.15$.  This bifurcation does not extend into the NIR in agreement with previous results \citep{Tegler2008}. It is interesting to note that the observed separation between the two groups is only $\sim0.1$ magnitudes, much smaller than that found in previous surveys. Rather, we observe a few moderately coloured Centaurs which have never before been detected in previous surveys of the Centaur population\citep{Tegler2008,Benecchi2011}.  

Unlike the blue Centaurs which exhibit a tight cluster, the red Centaurs span a broad swath of colours. This suggests that the red Centaurs exhibit a more diverse range of surfaces than do the blue Centaurs.

\begin{figure}[h] 
   \centering
      \epsscale{0.8}
   \plotone{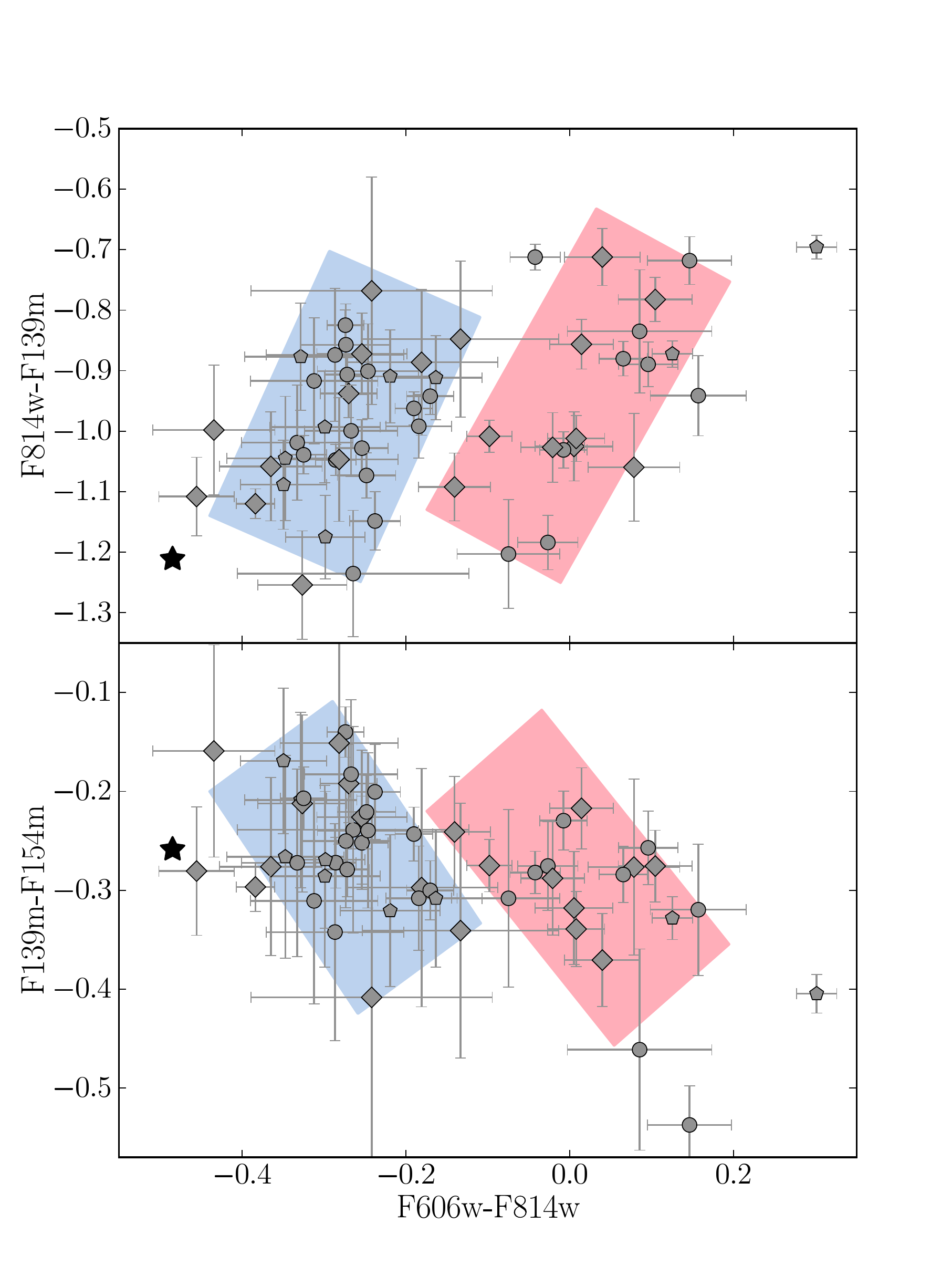} 
   \figcaption{(F814w-F139m) and (F139m-F154m) vs. (F606w-F814w) colours of the Centaurs (circles), scattered disk objects (pentagons), and resonant objects (diamonds) with $q<35$ AU. Solar colours are shown by the star. Only those objects with good measurements in all four filters are shown. The red and blue shaded regions show the approximately visual extent of the red and blue classes of low-$q$ object. The shaded regions and axis limits are the same as those shown in Figure~\ref{fig:CC_centaurs}. The low-$q$ objects consist of two separate classes of object.
   	\label{fig:CC_qlt35}}
\end{figure}

Analysis of ground-based colours suggests that the bifurcation exhibited by the Centaurs extends into the low perihelion scattered disk and resonant objects. This can be seen in Figure~\ref{fig:CC_q_MBOSS} where we present the (B-R) colours of objects with $q<35$~AU taken from the MBOSS dataset \citep{Hainaut2002}. The H/WTSOSS photometry support this finding. In Figure~\ref{fig:CC_qlt35} we present the observed colours of all objects with perihelia $q<35$ AU. This includes Centaurs and scattered, resonant, and hot classical objects. It can clearly be seen that most of the low perihelia objects, which include resonant and scattered disk objects, occupy nearly the same range of colours as the Centaurs, and exhibit the same bifurcation into two groups based on optical colour. Only two H/WTSOSS targets, Elatus (1999 UG5) and 1999 XX143 (121725), are found to have optical colours between that of the blue and red Centaur groups. The ground based colours of non-Centaur objects on the other hand have a higher fraction of objects with intermediate colours. These moderately coloured objects however, are predominately of objects with larger absolute magnitudes than in the Centaurs suggesting that the larger objects do not exhibit the colour bifurcation. Indeed, when only considering ground based observations of objects with $H\gtrsim6$ in the R band,  in the same range as the Centaur H/WTSOSS targets, the bifurcation is more pronounced suggesting that the largest objects occupy a different colour range than the smaller bodies.

A clear outlier, 2007 OR10 is the reddest and largest object in the H/WTSOSS sample, and is likely methane bearing \citep{Brown2011b}. As 2007 OR10 falls in a different class of object than the smaller methane-less objects, it is unsurprising that it appears as an outlier in Figure~\ref{fig:CC_qlt35}. 

When considered together, the blue, (F606w-F814w)$<-0.15$ low-$q$ objects appear as two distinct groups of objects each of which exhibits highly correlated optical and NIR colours. The Spearman rank correlation test \citep{Press2002} suggests that for the blue group the (F606w-F814w) colour correlates positively with the (F814w-F139m) colour and negatively with the (F139m-F154m) colour at the 99.8\% and 99\% significance levels respectively (see Figure~\ref{fig:CC_qlt35}). Despite the wider range of colours, like their blue counterparts, the red low-$q$ objects also exhibit  correlated optical and NIR colours which have a 99\% significance and correlate in a positive sense between (F606w-F814w) and (F814w-F139m) and in a negative sense between the (F606w-F814w) and (F139m-F154m) colours.


\subsubsection{Mixing Models}
A two component mixing model can readily explain the range of observed colours of each colour group of the low-$q$ objects. We demonstrate the effectiveness of a compositional mixture with the use of a simple Hapke surface model. Here we present a derivation showing only the necessary equations for our modelling. For a more thorough and complete derivation, see \citet{Hapke2002}.

We start with the  wavelength-dependent particle single scattering albedo $w\left(\lambda\right)$. In the two stream approximation, and under the assumption of irregular particles, the diffusive reflectance of the surface is given by

\begin{equation}
r_o = \frac{1-\gamma^*}{1+\gamma^*}
\end{equation}

\noindent
where $\gamma^*$ is the bulk albedo factor, and is given by

\begin{equation}
\gamma^* =\sqrt{1-w\left(\lambda\right)}.
\end{equation}

\noindent
This equation is valid for an infinitely thick layer. Then, for a spherical object, covered by isotropically scattering particles, and ignoring the opposition effect, the geometric albedo can be approximated by

\begin{equation}
A\left(\lambda\right) \sim 0.49 r_o + 0.196 r_o^2.
\label{eq:albedo}
\end{equation}

Consider a mixture model of two components $i$ and $j$. Each component will have a unique single scattering albedo, which we denote $w\left(\lambda\right)_i$ and $w\left(\lambda\right)_j$. These components can be mixed in two separate ways. The first, a geographic mixture, is one in which each component exists as discrete independent units on the surface. In this case, the effective albedo is just  the fractional sum of albedos of each component given by Equation~\ref{eq:albedo}, or

\begin{equation}
A\left(\lambda\right)_{geo} = f_{geo}A\left(\lambda\right)_i + (1-f_{geo})A\left(\lambda\right)_j.
\label{eq:albedo_geo}
\end{equation}

\noindent
where $f_{geo}$ is fraction of the surface occupied by component $i$.

Alternatively, the two components could exists as an intimate mixture, eg. mineral material in an icy matrix. In this case, the effective single scattering albedo is the fractional sum of the single scattering albedos of each component, and is given by

\begin{equation}
w\left(\lambda\right)_{int}=f_{int} w\left(\lambda\right)_i + (1-f_{int})w\left(\lambda\right)_j
\label{eq:w_int}
\end{equation}

\noindent
where $f_{int}$ is the fraction of the surface occupied by component $i$. Thus, with a choice in $w_i$ and $w_j$, Equation~\ref{eq:albedo_geo}, and \ref{eq:w_int} and \ref{eq:albedo} not only produce a range of colours, but also geometric albedos as well.

Analysis of Figure~\ref{fig:CC_qlt35} suggests that the correlated colours of the blue and red groups of low-$q$ objects might be explained by two separate mixing models that differ in their red components, but share a common neutral component whose colours are nearly Solar.  We tested this possibility by fitting both the geographic and intimate mixture models to the observations. We assume two separate branches of a mixing model of the same type (geographic or intimate). In our model, each branch has a different red component, and we assume that both mixtures share the same nearly neutral component.

The geographic mixture requires 9 parameters to describe all three observed colours simultaneously. These parameters are essentially the ratios of the single scattering albedos in each filter with respect to the F606w filter, three for each of the three components in our model. The intimate mixture requires two additional parameters describing the difference in single scattering albedos between each of the two red components and the neutral component. A change in these additional parameters results in a change in curvature of the resultant colour curve. 

The three component geographic and intimate mixture models were fit to the observations in a least-squares sense. We chose (F606w-F814w)=-0.15 to separate the objects belonging to the blue and red groups during our fits. As well, it was necessary to ensure that the resultant range of model colours was restricted to the range of observed colours of our targets. This action was taken to prevent run-away during the least-squares minimization.  It should be emphasized that it cannot be determined if the true colours are beyond this range by the observed colours alone.

\begin{figure}[h] 
   \centering
      \epsscale{0.8}
   \plotone{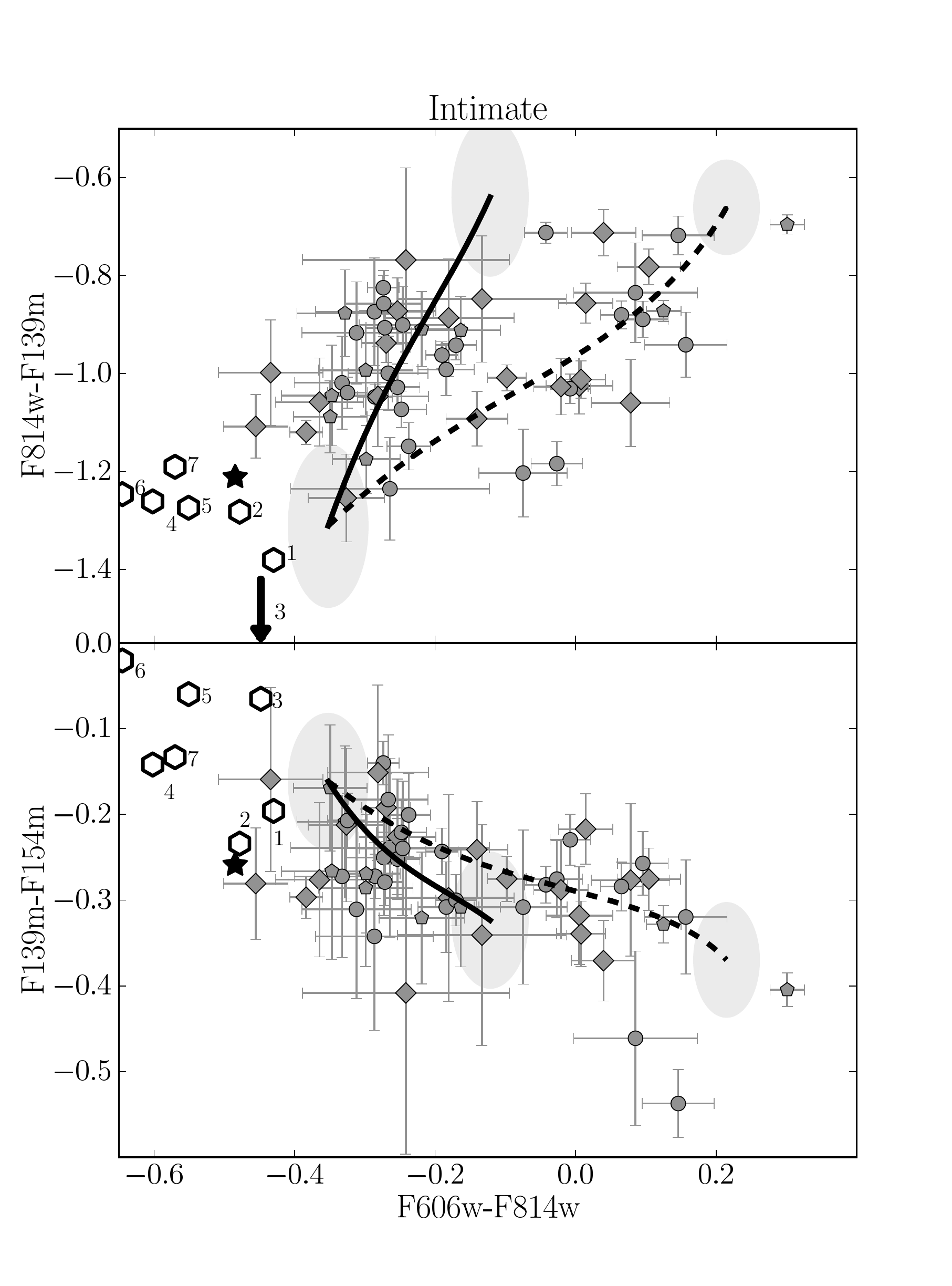} 
   \figcaption{As in Figure~\ref{fig:CC_qlt35} but with the colours predicted by the red and blue branches of the intimate mixture model shown as dashed and solid lines respectively. The 3-sigma uncertainty range in the colours of both red and the neutral-end components are shown as shaded regions. The colours of a selected range of silicates consistent with the inferred colour of the neutral component are shown as open hexagons. 1, 2 - chlorites CU91-238A, and GDS159. 3, 4 - serpentines HS318.4B and HS8.3B. 5, 6 - olivines GDS70.d and NMNH137044.b. 7 - magnetite HS195.3B. The intimate mixture models fully describe the observed KBO colours. \label{fig:CC_qlt35_wModel_intimate}}
\end{figure}

\begin{figure}[h] 
   \centering
      \epsscale{0.8}
   \plotone{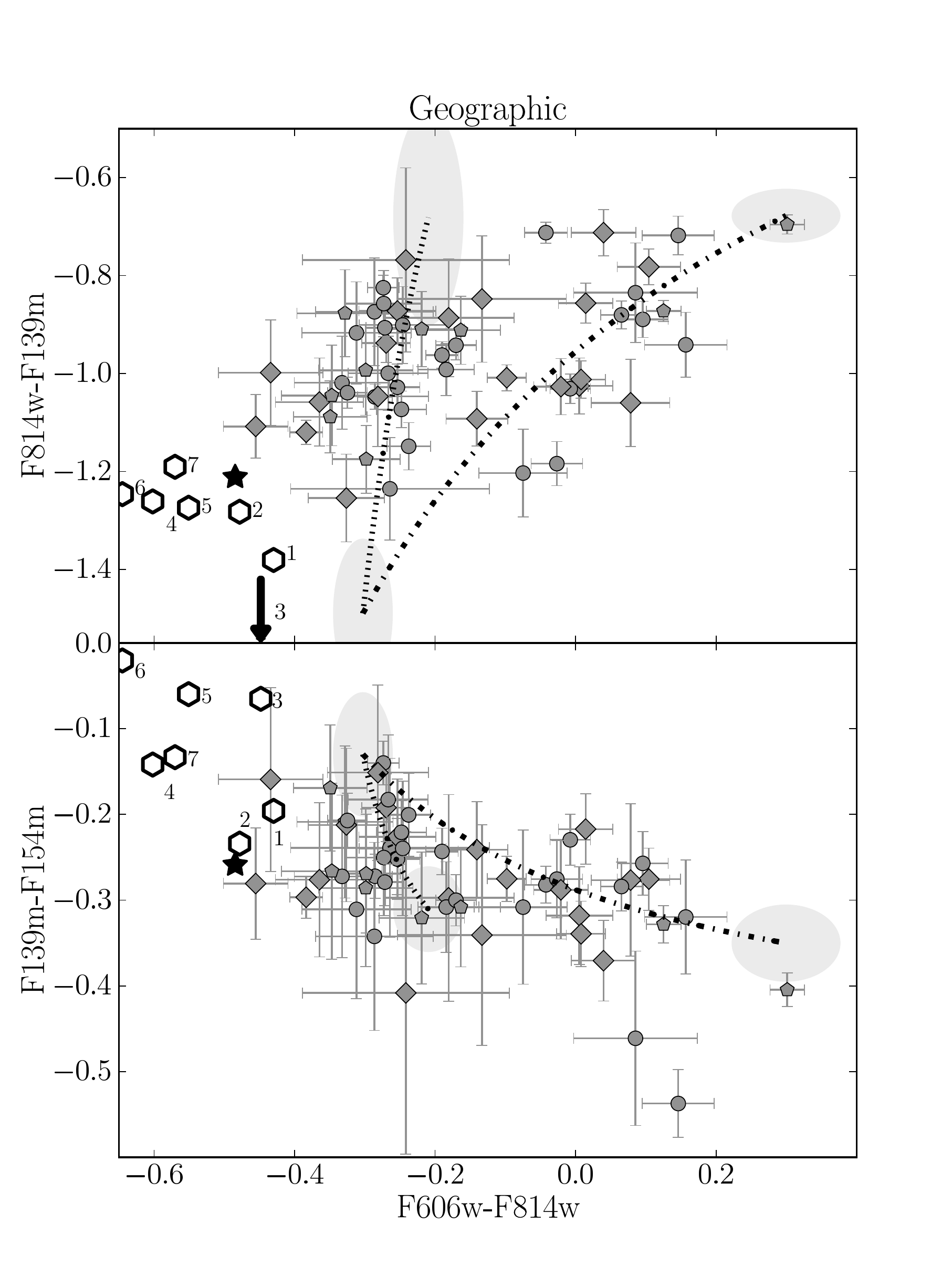} 
   \figcaption{As in Figure~\ref{fig:CC_qlt35_wModel_intimate} but with the colours predicted by the red and blue branches of the geographic mixture model shown as dash-dotted and dotted lines respectively. The 3-sigma uncertainty range in the colours of both red and the neutral-end components are shown as shaded regions. The colours of a selected range of silicates consistent with the inferred colour of the neutral component are shown as open hexagons. 1, 2 - chlorites CU91-238A, and GDS159. 3, 4 - serpentines HS318.4B and HS8.3B. 5, 6 - olivines GDS70.d and NMNH137044.b. 7 - magnetite HS195.3B. Like the intimate mixture models, the geographic models well describe the observed KBO colours, but with higher residuals and one additional outlier. \label{fig:CC_qlt35_wModel_geographic}}
\end{figure}

The results of the model fitting to the 55 low-$q$ objects in the H/WTSOSS sample are shown in Figures~\ref{fig:CC_qlt35_wModel_intimate} and \ref{fig:CC_qlt35_wModel_geographic}. As can be seen, the colours of the neutral and red components for both models are very similar. Both the intimate and geographic mixture models do an adequate job of describing the observations. 

2007 JG43, 2004 EW95, and 2004 TV357, all blue low-$q$ objects , along with red low-$q$ objects, Elatus (1999 UG5), Crantor (2002 GO9) and 2007 OR10 were all identified as outliers responsible for the vast majority of the fit residuals. Crantor has optical colour very near, but just red-ward of the (F606w-F814w)=-0.15 colour chosen to separate the blue and red Centaurs. Given its NIR colours however, Crantor is roughly equally consistent with either the red or blue branches of the mixing model. As discussed above, 2007 OR10 possesses a surface unique to the large volatile bearing KBOs, and it is therefore no surprise that it stands out from the majority. It is not unreasonable to expect other outliers as our simple mixing models do not consider other processes such as collisions which can alter surface colours. When the outliers are not considered in the fit, the resultant chi-squared values were reduced by nearly half compared to when the outliers were considered.
 
 A similar finding as for Crantor exists for a few of the blue low-$q$ objects. That is, objects placed in the blue group by our selected division are better described by the blue-end of the red mixing curve. It is not clear however, if this is a result of the limitations of our chi-squared fits. This possibility should be further studied with more observations however, as the result would imply that the red Centaur population is not uniquely red, but rather occupies a range of optical colours that sparsely samples the blue end of the red mixing curve.

After excluding the 6 outliers, the model best-fits had reduced chi-squared values of $\chi_{\nu,geo}^2=2.87$ and $\chi_{\nu,int}^2=3.13$ with 38 and 40 degrees of freedom for each model respectively. The geographic model only resulted in a reduction of the variance in the data of only 13\% while a reduction of 40\% was found for the intimate mixture model. In addition, though the best-fit intimate mixture could account for the colours of 2006 QP160, the best-fit geographic model could not and rejected it as an additional outlier. Despite the two extra parameters, the intimate mixture model provides a better description of the data over the geographic model. These results suggest that intimate mixtures are a better model of the surfaces of the objects we observed compared to the geographic mixtures.

Typically, inference of KBO compositional make-up has been made through extensive spectral modelling of not only deep spectral lines diagnostic of particular materials, but also of the general spectroscopic shape not attributable to any one material \citep[see][for a review]{Barucci2008}. Materials required to successfully model  featureless spectra of KBOs typically include organic materials such as tholins or other irradiated hydrocarbons, mineral components such as olivine, water-ice, and neutral darkening agents such as carbon  \citep{Cruikshank1998,Alvarez-Candal2008,Merlin2010,Barucci2011}. The range of colours required by the components of our mixing model are compatible with many of these materials.

The red components require a material which is blue from $\sim1.3$ to $1.5 \mbox{ $\mu$m}$. It is tempting to attribute this colour to water-ice absorption. The red colours of each material for wavelengths $\lesssim1.3$ are compatible with the irradiated organic ices and tholins often attributed to the spectral shapes of other KBOs \citep{Cruikshank1991,Brunetto2006}. While no unique identification can be made from these observations, it seems plausible that the red components are different combinations of organic and water ices.

\begin{figure}[h] 
   \centering
   \plotone{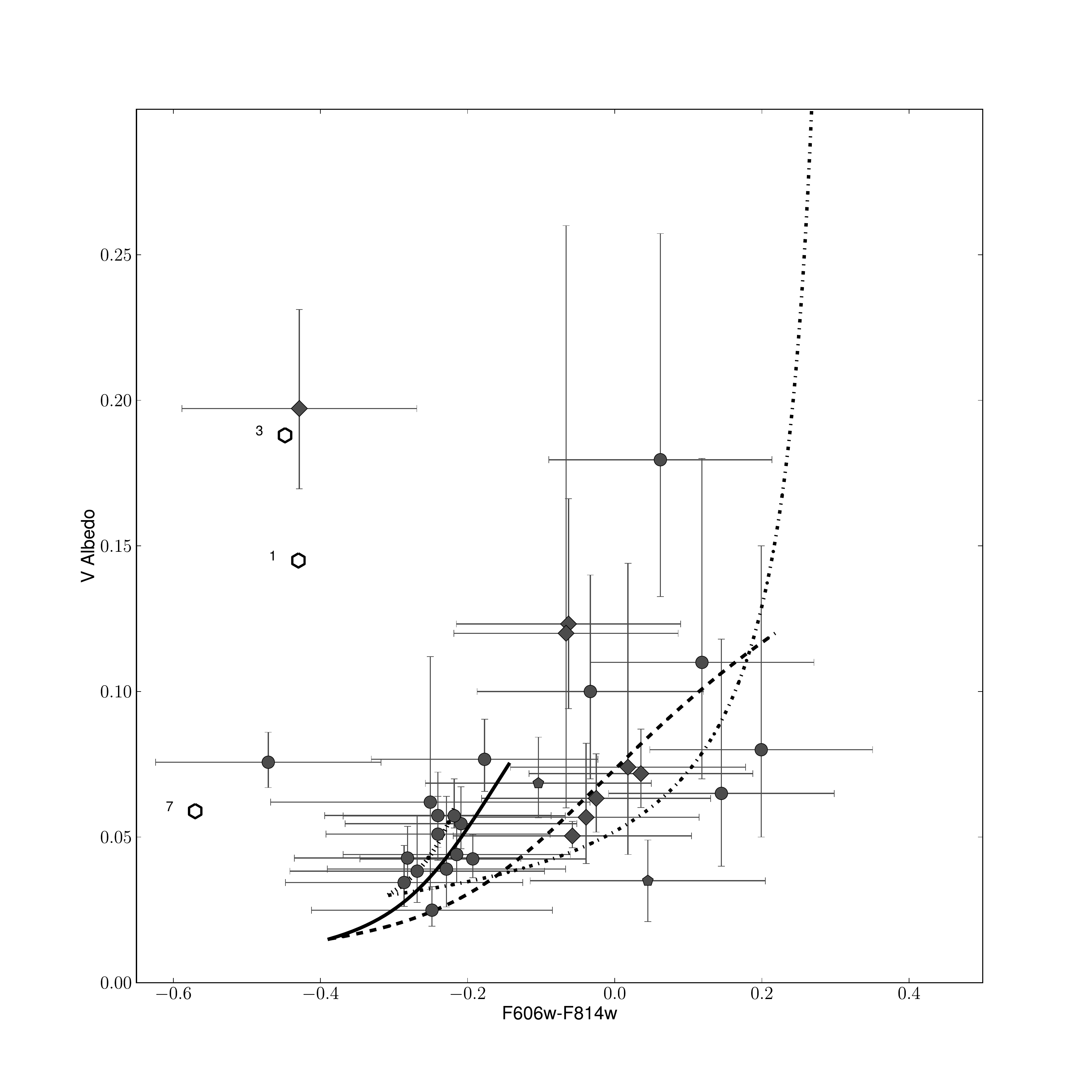} 
   \figcaption{Geometric visible albedo vs. (F606w-F814w) colour calculated from available ground-based optical colours from the MBOSS dataset \citep{Hainaut2002} for small KBOs with $q<35$ AU. Symbols are as in Figures~\ref{fig:CC_qlt35}, \ref{fig:CC_qgt35} and \ref{fig:CC_cc}. The colours and albedos predicted by red and blue intimate and red and blue geographic mixture models which describe the low-$q$ objects are shown as solid and dashed dotted and dash-dotted lines.  The conversion from ground-based to HST filters was done using the Solar analog colours produced by the {\it synphot iraf} routine which results in a $\sim0.15$ magnitude uncertainty. The albedos of the silicates from Figure~\ref{fig:CC_qlt35_wModel_intimate} are shown as hexagons. Those not visible have albedos higher than the range shown. \label{fig:alb_qlt35}}
\end{figure}

The range of colours of the neutral component require a material that is virtually neutral across all four filters used in our observations. Due to the large absorption at $\sim 1.55 \mbox{ $\mu$m}$ water-ice is not a candidate. Rather, a plausible candidate for the neutral component appear to be silicates. The colours of various minerals taken from the USGS spectral library \footnote{Clark, R.N., Swayze, G.A., Wise, R., Livo, E., Hoefen, T., Kokaly, R., Sutley, S.J., 2007, USGS digital spectral library splib06a: U.S. Geological Survey, Digital Data Series 231, \url{http://speclab.cr.usgs.gov/spectral.lib06}} are shown in Figures~\ref{fig:CC_qlt35_wModel_intimate} and \ref{fig:CC_qlt35_wModel_geographic} along with their albedos in Figure~\ref{fig:alb_qlt35}. Minerals consistent in colour with the inferred neutral component  include olivines similar to those inferred from previous spectral modelling of KBOs \citep{Merlin2010}. Other compatible minerals include serpentines and aqueously altered phyllosilicates, the latter of which provides the best match to our fitted colours. The albedos of all of these silicate materials however, are all too high to be consistent with the low $\rho\sim5\%$ albedos of the blue Centaurs. Rather some additional neutral darkening agent such as amorphous carbon is needed, as has been found in previous spectral modelling efforts \citep[see for instance][]{Cruikshank1998}.

The candidate silicate materials we and others have presented are similar to the iron rich minerals and hydrated silicates that have been suggested to exist on the largest, outermost Saturnian irregular satellite, Phoebe \citep{Clark2005}. Phoebe and the other irregular satellites are captured objects which likely originated from the same primordial population as the observed KBOs \citep{Nesvorny2007}. Thus, while no claim from our observations can be made as what the neutral component may be, the apparent compositional similarity between Phoebe and the observed KBOs suggests that indeed these objects share a genetic link, and that the minerals we present are good candidates for the neutral material. If confirmed, this result would suggest that despite the large formation distances of KBOs and irregular satellites, aqueous alteration and therefore liquid water was prevalent in the outer Solar system.

The best-fits of the intimate mixture to the colours of the blue objects have a preference to models in which the  red component shares a similar single scattering albedo to the blue component. The red objects however, are better matched if the red component albedo of the red branch is significantly larger than that of the neutral component. Despite the poor availability of albedo measurements, many of which are thermal model dependent, our findings are in agreement with observed trend in KBO albedo with colour. We present measured albedos of KBOs \citet[see][]{Stansberry2008} versus their F606w-F814w colour calculated from their observed colours taken from the MBOSS dataset in Figure~\ref{fig:alb_qlt35}. This suggests that the reddest objects have higher albedos than the bluest objects. The observed colours and albedos are well  reproduced by the intimate mixture model if the red components of the blue and red branches have albedos a factor of $\sim2$ and $\sim4$ times larger than that of the neutral component (assumed to be $3$\% in Figure~\ref{fig:alb_qlt35}). The geographic mixture model can also reproduce the albedo colour trend, but predicts that the reddest objects will have albedos $\gtrsim25$\%, a factor of $\sim3$ higher than that predicted for the intimate model. Further observations are required to determine if such high albedos exist for reddest KBOs.

The chi-squared fits of the mixture models we present here have a few potential weaknesses. As stated above, the colours of the two components were restricted to prevent run-away during the minimization. As such, the true colours of the neutral and red components might be beyond the range of colours inferred from the fits. In addition, the model we present is rather simple, and does not account for many processes which may affect the surfaces of KBOs. As such, the fits we present are best used as a guide to understanding the primary surface properties of objects in the Kuiper belt rather than an exact description. Despite the model's simplicity, it is clear that two separate branches of a mixing model which shares a common - and neutral - component can account for the major trends in the observed colours of the low-perihelion KBOs. In addition, a model in which the components are mixed intimately appears to be favoured over one in which the mixture is geographic.

\subsection{The High Perihelion Objects \label{sec:high_q}}
\begin{figure}[h] 
   \centering
      \epsscale{0.8}
   \plotone{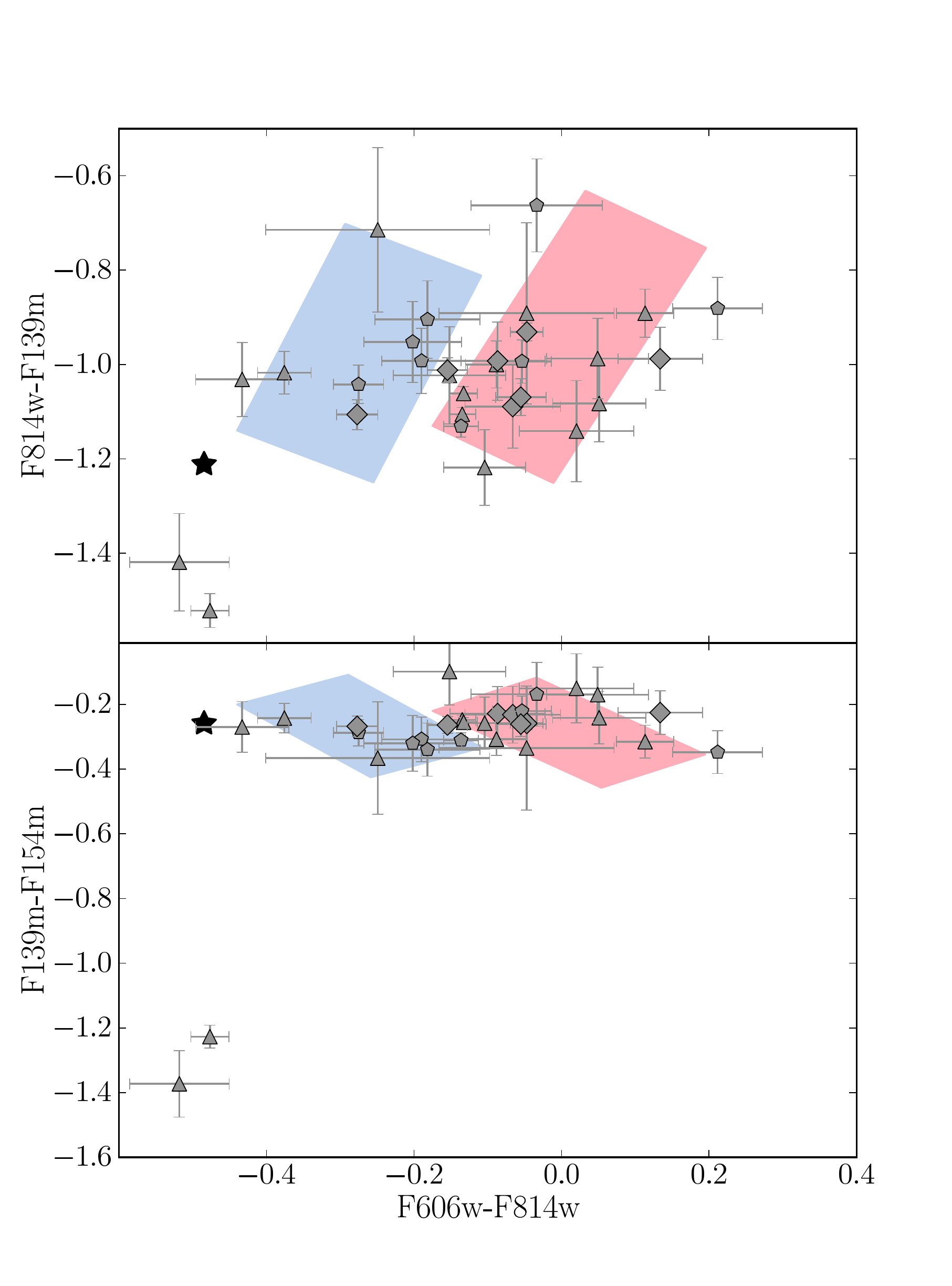} 
   \figcaption{(F814w-F139m) and (F139m-F154m) vs. (F606w-F814w) colours of the Centaurs (black circles), scattered disk objects (pentagons), hot classical objects (diamonds), and resonant objects (diamonds) with $q>35$ AU. Haumea family members 2003 UZ117 and 2003 SQ317 stand out as having very strong water-ice absorption. Solar colours are shown by the star. Only those objects with good measurements in all four filters are shown. The red and blue shaded regions approximately show the extent of the red and blue classes and are the same as those shown in Figure~\ref{fig:CC_centaurs}.  \label{fig:CC_qgt35}}
\end{figure}

The observed colours of the hot classical, scattered and resonant objects with $q>35$ AU are shown in Figure~\ref{fig:CC_qgt35}. Other than the two identified Haumea family members, 2003 UZ117 and 2003 SQ317 \citep{Schaller2008,Snodgrass2010}, many of the distant excited population are consistent in colour with the intimate mixture model fit to the low-$q$ population. The distant population however, has a higher fraction of objects with moderate optical colour, $-0.2\lesssim$(F606w-F814w)$\lesssim 0.0$ and suggests that the colours of the distant objects are not consistent with the intimate mixture model.  No obvious hints of a bimodal population are apparent for the high-$q$ objects.

We tested whether or not the non-Haumea family, high-$q$ objects could be drawn from the mixture model of the low-$q$ objects with a monte-carlo technique. This involved drawing from the dual intimate mixture model, a sample of objects equal in number to the high-$q$ population with the same ratio of red to blue objects. Gaussian scatter was added to the random colours with standard deviations equal to the observed uncertainties. Then, by adopting the observed uncertainty distribution, the intimate mixing model that was used to describe the low-$q$ objects was refit to the random sample. This process was repeated 100 times, and the distribution of chi-squared values was recorded. In 90\% of cases, the simulated chi-squared values were lower than the observed value demonstrating that the model which describes the low-$q$ objects does not match the observed high-$q$ population. 

As discussed above for 2007 OR10, the largest objects have a different colour distribution than smaller objects. Using the observed absolute magnitude as a proxy for size, the observed high-$q$ population contains objects with absolute magnitudes $4\lesssim H_{606}\lesssim 8$, while the low-$q$ population, ignoring 2007 OR10, only has objects with $H_{606}>6$. Indeed, the largest objects in our sample are the most discrepant. After rejecting the 8 intermediate sized objects with high-$q$ objects with $H_{606}<5.6$,  the monte-carlo test suggests that the probability of drawing a random chi-squared equal to or greater than the observed value for the remaining 20 objects is 35\%; the colours of the small, high-$q$ objects are consistent with being drawn from the intimate mixture model that describes the colours of the low-$q$ objects. It should be noted that object 2001 CZ31 appears as an outlier. Indeed when this object is rejected, the probability of drawing a random chi-squared equal to or greater than observed value becomes 50\%. Though rejection of this object as an outlier is not formally required by the model.

\begin{figure}[h] 
   \centering
      \epsscale{0.8}
   \plotone{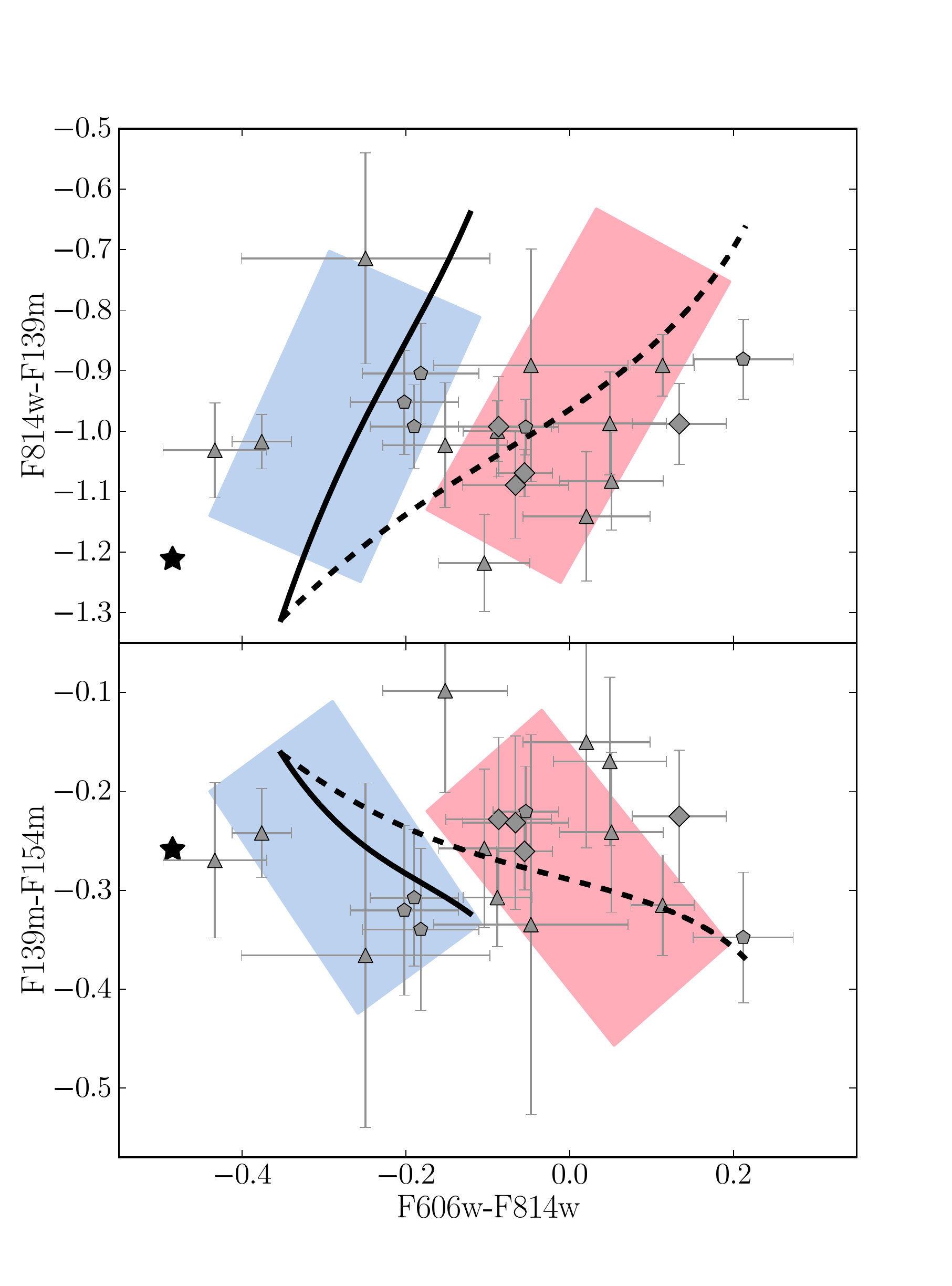} 
   \figcaption{As in Figure~\ref{fig:CC_qgt35} but only for those non-Haumea family member objects with $H_{606}>5.6$. The blue and red intimate mixture models which describe the colours of the low-$q$ objects are shown as solid and dashed lines respectively. The red and blue shaded regions approximately show the extent of the red and blue classes of low-$q$ object. The shaded regions and axis limits are the same as those shown in Figure~\ref{fig:CC_centaurs}. The small, excited objects are consistent with being drawn entirely from the red and blue classes of low-$q$ object.\label{fig:CC_qgt35_wModel}}
\end{figure}

The colours of the small, excited, high-$q$ population are shown in Figure~\ref{fig:CC_qgt35_wModel}. No bifurcation of the optical colours is apparent in this figure. It should be noted that this sample has uncertainties which are $\gtrsim50\%$ larger than the low-$q$ sample. Thus, we should not expect to see the bimodality exhibited by the low-$q$ objects. The observations we present here are not of sufficient quality to show the existence of such a bimodality. The colours of the small, excited, high-$q$ objects however, are consistent with the bimodal low-$q$ population.

Our results suggest that only the largest objects in our high-$q$ sample are inconsistent with the low-$q$ population. The larger uncertainties could mask small changes in colour between the two samples. If it is true that the small high-$q$ objects posses the same colour distribution as the low-$q$ objects, this would suggest that little to no colour evolution occurs when high-$q$ objects are scattered into the Centaur region. Even if colour evolution does occur, as typical colour uncertainties are only of order 5\%, the changes must be small.

The largest objects in our sample have colours inconsistent with the two intimate mixtures which describe the low-$q$ population. As these objects must have formed of the same material as the smaller objects, it must be that some processes which do not readily occur on smaller objects have altered the surfaces of the larger bodies. Our observations suggest that the size transition occurs at $H_{606}\sim5.6$. Assuming a 6\% albedo, this corresponds to objects with diameters, $D\sim400$km. This also coincides with a transition in albedos; objects smaller than this size have lower albedos than larger objects. The larger, intermediate sized objects are still too small to retain their primordial volatile budget as the largest known KBOs have. But for these intermediate sized objects, effects such as the onset of differentiation could modify their surfaces. What ever the cause might be, the apparent difference in colours between objects with absolute magnitudes smaller and larger than $H_{606}\sim5.6$ clearly warrants confirmation.

\begin{figure}[h] 
   \centering
      \epsscale{0.8}
   \plotone{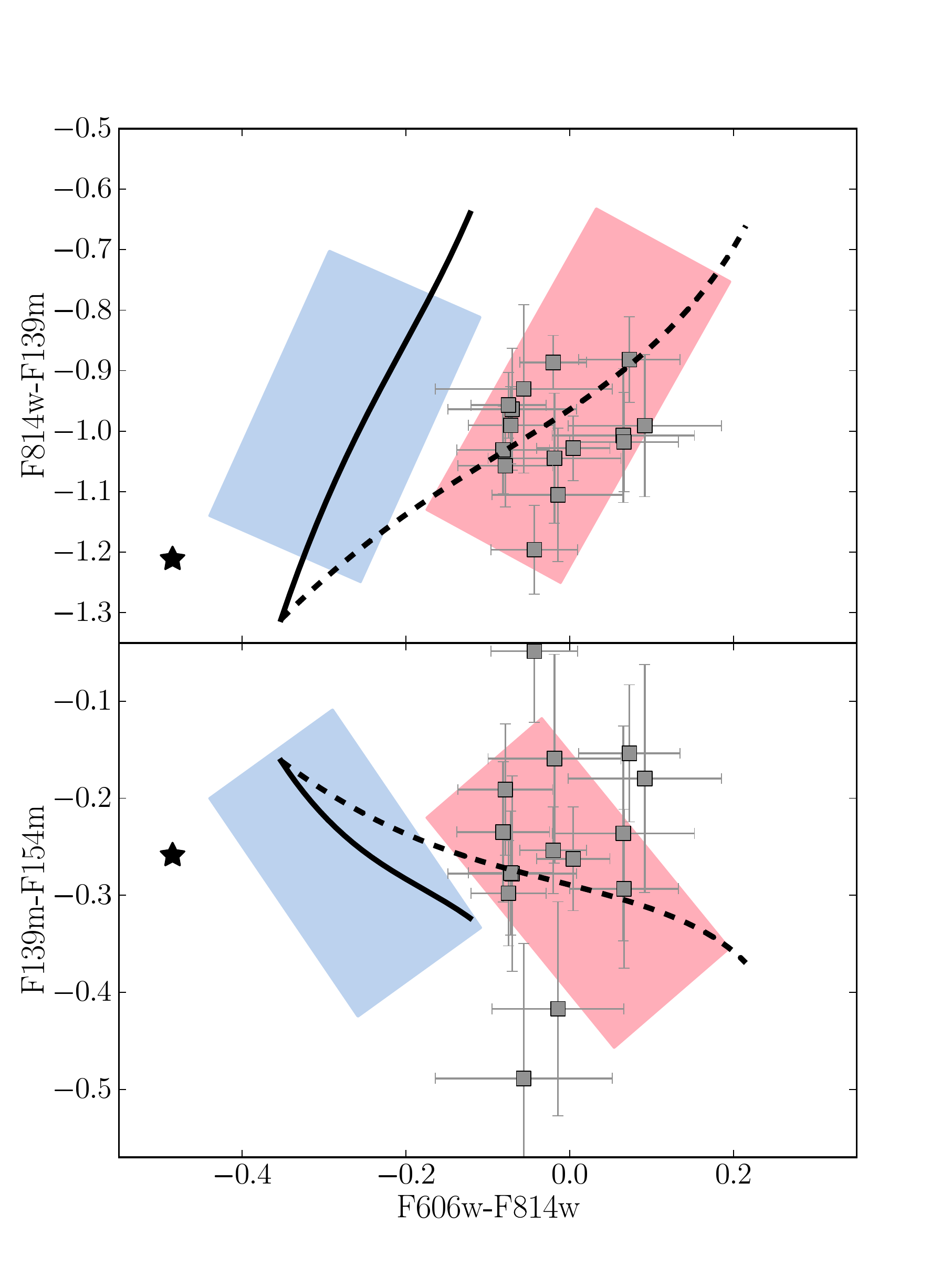} 
   \figcaption{(F814w-F139m) and (F139m-F154m) vs. (F606w-F814w) colours of the cold classical objects, shown as red squares. Solar colours are shown by the star. Only those objects with good measurements in all four filters are shown. The red and blue intimate mixture models which describe the colours of the low-$q$ objects are shown as dashed and solid lines respectively. The red and blue shaded regions approximately show the extent of the red and blue classes of low-$q$ object. The shaded regions and axis limits are the same as those shown in Figure~\ref{fig:CC_centaurs}. The cold classical objects do not appear to be consistent with either class of excited object.
    \label{fig:CC_cc}}
\end{figure}

The observed colours of the cold classical objects are presented in Figure~\ref{fig:CC_cc}. The well known unique colours of this population are clear, lacking not only the reddest objects, but also the entire blue group of objects observed in other KBO populations. Interestingly, with the exception of object 2003 HG57, the CCOs appear consistent in colour with the red branch of the intimate mixture model that describes the low-$q$ population, albeit with a narrower range of colours. Indeed, our monte-carlo tests suggest the colours of the CCOs are entirely consistent with the red-branch of the mixture model; the probability of a randomly drawn chi-squared value greater than that observed is only 20\% when 2003 HG57 is excluded.

Our observations alone would suggest that the CCOs and the red branch of the Centaurs and small excited KBOs share different mixtures of the same primordial materials. This simple picture is challenged by available albedo observations of these objects. Despite the sparser sampling and larger error-bars, \citet{Brucker2009} has shown that the albedos of the CCOs $\gtrsim0.15$, inconsistent with that predicted from the red-branch of the mixing model. The excited high-$q$ objects  however, are fully consistent with the model and exhibit the same general trends in albedo as the low-$q$ objects; the red objects appear to have higher, and exhibit a wider range of albedos than do the blue objects which all appear to have low albedos of $\sim 5\%$. Thus, while it appears that the small, excited populations of the Kuiper belt are all drawn from either the red or blue branches of the mixing model, the CCOs are not. Higher quality albedo measurements are required to determine whether or not the cold classical objects posses a surface unique to this population alone.

\subsection{Bifurcation Tests}
In Section~\ref{sec:low_q} we identified an apparent bifurcation in the colours of the low-$q$ objects which is consistent with that observed in the ground-based optical colours of the Centaurs \citep{Tegler2008}. While the colours of all small, excited objects appear to exhibit this bifurcation, the significance of this feature, and the physical model we put forth to explain it remains untested. We turn to the use of ``blind'' statistical tests. That is, tests which require no a priori information or user input to test if the data exhibit two separate populations, or are consistent with being drawn from a single population. The first test we turn to is the non-parametric bifurcation test commonly adopted in the literature pertaining to the colours of KBOs, Hartigan's DIP test \citep{Hartigan1985}. The DIP test however, is only 1-dimensional. Thus, when utilizing the DIP test, we must only consider the optical colour alone where the data appear most bifurcated. For the multivariate data we present here the 1-dimensional DIP test is inappropriate.

We also consider multivariate clustering techniques. The first clustering technique we consider is hierarchical clustering using Ward's criterion \citep{Everitt2011}. Ward's criterion which uses a sum of squares of cluster member distances from average cluster position as a clustering metric, and typically results in same sized, circular clusters. In addition, we consider the normal mixture model distribution clustering technique which assumes underlying multivariate Gaussian distributions for each of the identified clusters. In a Bayesian formalism, this technique fits a chosen number of Gaussian components or clusters to the observed distribution resulting in a best-fit likelihood, $L$. As such, the corrected Bayes Information Criterion (BIC)  which can be approximated by $-2\log{L}+p\log{\frac{n-2}{24}}$ can be evaluated for different cluster numbers to provide subjective information as to the true number of clusters within the data. Here $p$ is the number of parameters in the fit (12 per Gaussian component in 3D) and $n$ is the number of data points. That is, the number of clusters can be increased one at a time such that the resultant increase in the BIC is maximized. This provides the most likely number of clusters. Mixture model maximum likelihood fits to the data were performed with the {\it PyMix} package \citep{Georgi2010}. A review of these and other common clustering techniques is available in \citet{Everitt2011}.

We also consider the minimum spanning tree based technique, \textbf{MSDR}, presented by \citet{Grygorash2006}. This technique is excellent at identifying clusters in small sample, multivariate datasets and avoids the chaining effect which other hierarchical and minimum spanning tree based techniques are prone \citep{Everitt2011}

Much like virtually all other clustering methods however, the three clustering techniques mentioned above, hierarchical, mixture model, and \textbf{MSDR}  provide virtually no information as to the correct number of clusters in the observed data. When this information is available, it is subjective at best - no test of the significance of identified clusters is provided. To overcome this weakness, we introduce the {\it F optimal plane} or \FOP. The \FOP  which is also based on the minimum spanning tree of the data, uses a  separate clustering statistic than  \textbf{MSDR} to determine which edge to prune resulting in two sub clusters of the data. In addition, the resultant statistic $F$ of the pruned edge provides a means of determining the significance that the two sub clusters are consistent with the null hypothesis, that the data are consistent with a single population only. Thus, the \FOP has a major advantage over other multivariate clustering techniques in that it provides quantitative information as to the number of extant clusters in the observed data.

A detailed discussion of the \FOP and some examples of its performance versus the \textbf{MSDR} clustering technique are presented in the appendix. The reader is warned that the \FOP is an ad-hoc combination of two alternative clustering metrics that remains to be mathematically proven. As is shown in the appendix however, the \FOP performs just as well and often even outperforms the mathematically tested \textbf{MSDR} based clustering technique in both correctly identifying the true number of clusters as well as correctly assigning cluster membership in various simulated datasets.

\begin{figure}[h] 
   \centering
      \epsscale{0.8}
   \plotone{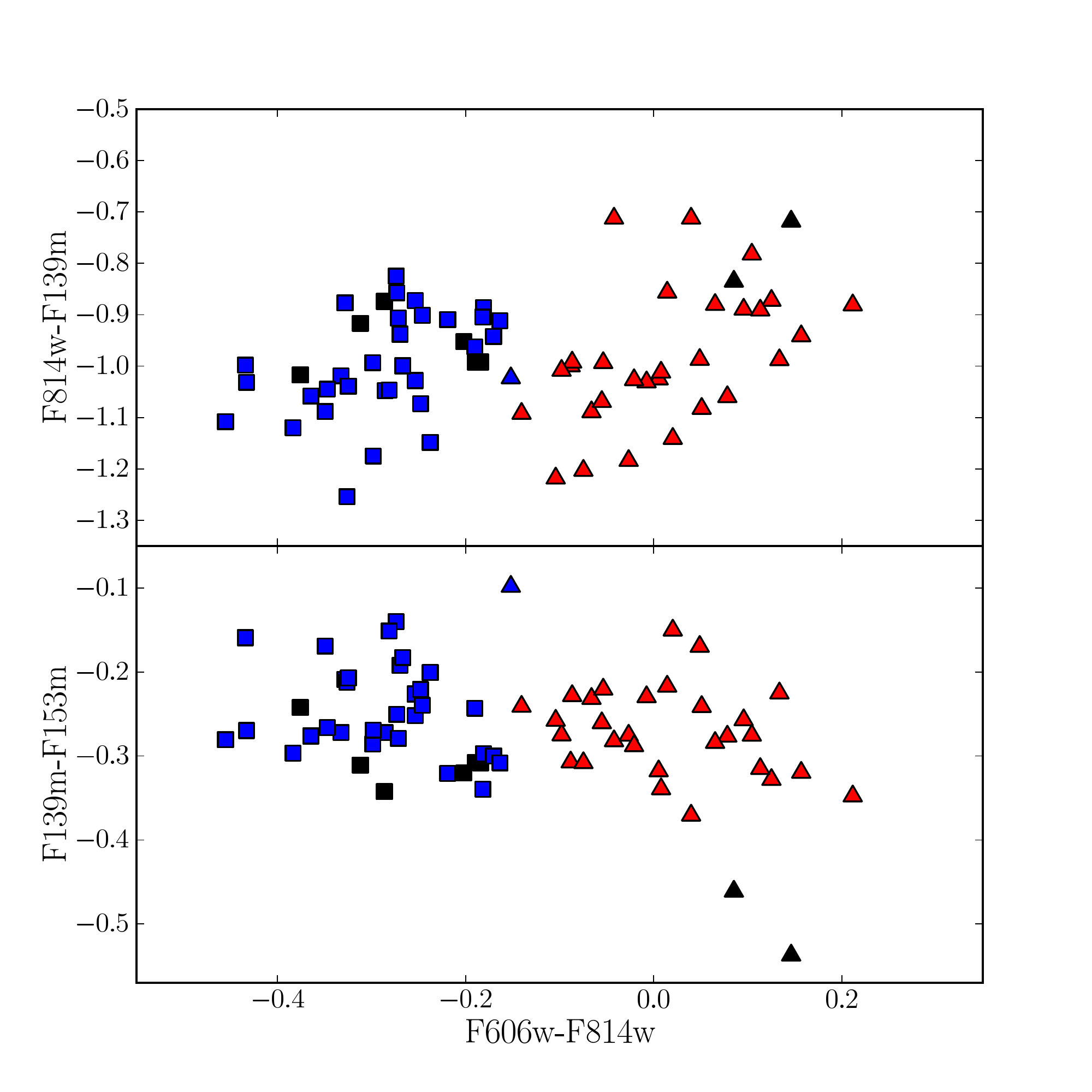} 
   \figcaption{The results of the hierarchical clustering and normal mixture model clustering applied to the (F814w-F139m) and (F139m-F154m) vs. (F606w-F814w) colours of non-CCO, non-family member objects with $H>5.6$ and error in (F814w-F139m) less than 0.1 mags. Errorbars are omitted for clarity. Squares and triangles are objects classified as belonging to the two separate clusters identified by the hierarchical clustering. Blue, red and black points are objects classified as belonging to the three separate clusters identified by the normal mixture model clustering. Both techniques agree with our assertion that most of the small, excited KBOs exhibit bifurcate into two separate classes at a colour of (F606w-F814w)$\sim-0.15$.
 \label{fig:HMM_excited}}
\end{figure}

\begin{figure}[h] 
   \centering
      \epsscale{0.8}
   \plotone{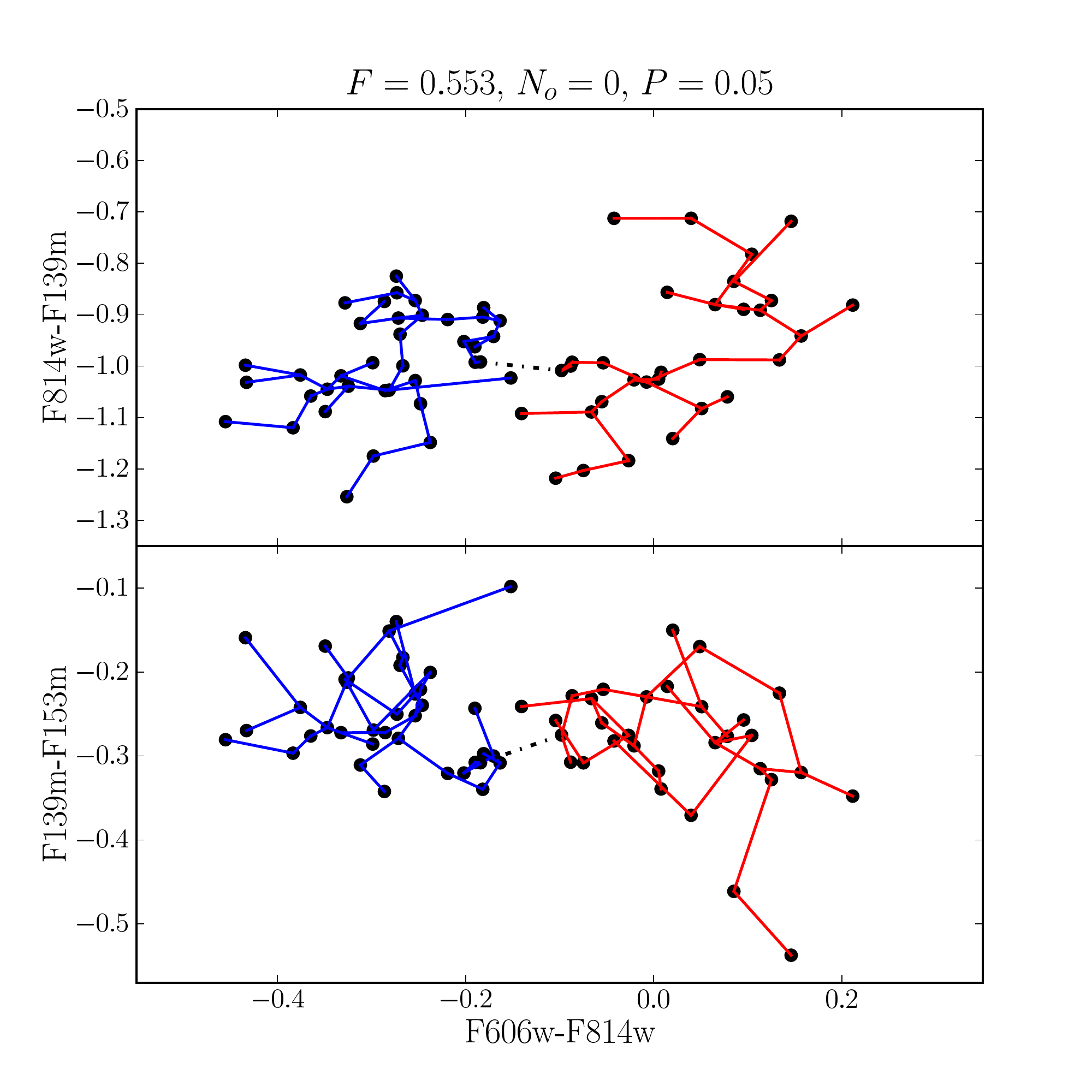} 
   \figcaption{The results of the \FOP and \textbf{MSDR} clustering to the (F814w-F139m) and (F139m-F154m) vs. (F606w-F814w) colours of non-CCO, non-family member objects with $H>5.6$ and error in (F814w-F139m) less than 0.1 mags. Errorbars are omitted for clarity. The edges of the minimum spanning tree are shown as line segments. The dotted segment shows the pruned edge identified by both techniques, resulting in the two clusters identified by the blue and red segments. The $F$, $N_o$  and the probability that the sample consists of only one population from the \FOP are shown. Both techniques agree with our assertion that most of the small, excited KBOs exhibit bifurcate into two separate classes at a colour of (F606w-F814w)$\sim-0.15$.
 \label{fig:ST_excited}}
\end{figure}

In Figures~\ref{fig:HMM_excited} and \ref{fig:ST_excited} we present the results of the blind statistical tests applied to the observed colours of the small, excited population, or those non-CCO, non-family member objects with $H>5.6$. For this test, we further reject the data with error in (F6062-F814w) of 0.1 mags or larger to avoid biasing our results with poor quality data. 

In the (F606w-F814w) colour alone, the small, excited sample appears bimodal; the DIP test confirms this with 98.6\% chance of bimodality. The application of the hierarchical clustering technique with two clusters results in two clusters which exactly reflect the red and blue compositional classes identified in Sections~\ref{sec:low_q} and \ref{sec:high_q}. That is, those blue class members are all uniquely identified to one cluster, and those red class members are uniquely identified to the other cluster with the no incorrect identifications.

Application of the normal mixture models paints a similar picture. The value of the BIC as the number of clusters is increased from 1 to 5 is a,b,c,d,e with a maximal increase when the number of clusters is three. The first two identified clusters are just the red and blue compositional classes. Most of the red objects fall into the first cluster, and most of the blue objects fall into the blue cluster, with only 8 objects (6 red and 2 blue) falling into the third cluster which occupies nearly the full extent of the blue and red clusters combined. Thus, the normal mixture model clustering paints the same picture as we discuss above; the small, excited KBOs appear to exhibit to compositional classes with a few outlier objects.

The results of the \FOP are in agreement with the hierarchical and normal mixture model results. The \FOP identifies two clusters within the data - exactly the two classes we identified in Sections~\ref{sec:low_q} and \ref{sec:high_q}. The edge pruned by this method is the only edge of the minimum spanning tree that spans (F606w-F814w)=-0.15, the colour which divides the two compositional classes. The results of the \FOP agree; calibrating the result of the \FOP by bootstrapping and by sampling from the uniform distribution demonstrate that the probability that the observed data are consistent with a single population is only 5\% or 7\% respectively (see the Appendix for a discussion of these calibration methods).

The \textbf{MSDR} technique further strengthens this result. The first two sub-clusters found by this technique are those visually identified in Sections~\ref{sec:low_q} and \ref{sec:high_q} and are identical to those found by the \FOP. The stopping criterion of the \textbf{MSDR} technique is met when 4 clusters are produced. The \FOP however, demonstrates that further divisions beyond the first are insignificant, and only two populations are required.

Upon application of these blind tests to the Centaur and low-$q$ populations, we find the same results, albeit with lower significance. Hartigan's DIP test suggests that the probability of bimodality is $85$ and $67$\% for the Centaurs and low-$q$ samples respectively. The \textbf{MSDR} clustering and \FOP both identify identical sub-populations, the same two classes discussed above separated by (F606w-F814w)=-0.15. For the Centaur and low-$q$ subsets, the probability that the sample consists of a single population is $14$ and $30$\% respectively.  

\begin{figure}[h] 
   \centering
      \plotone{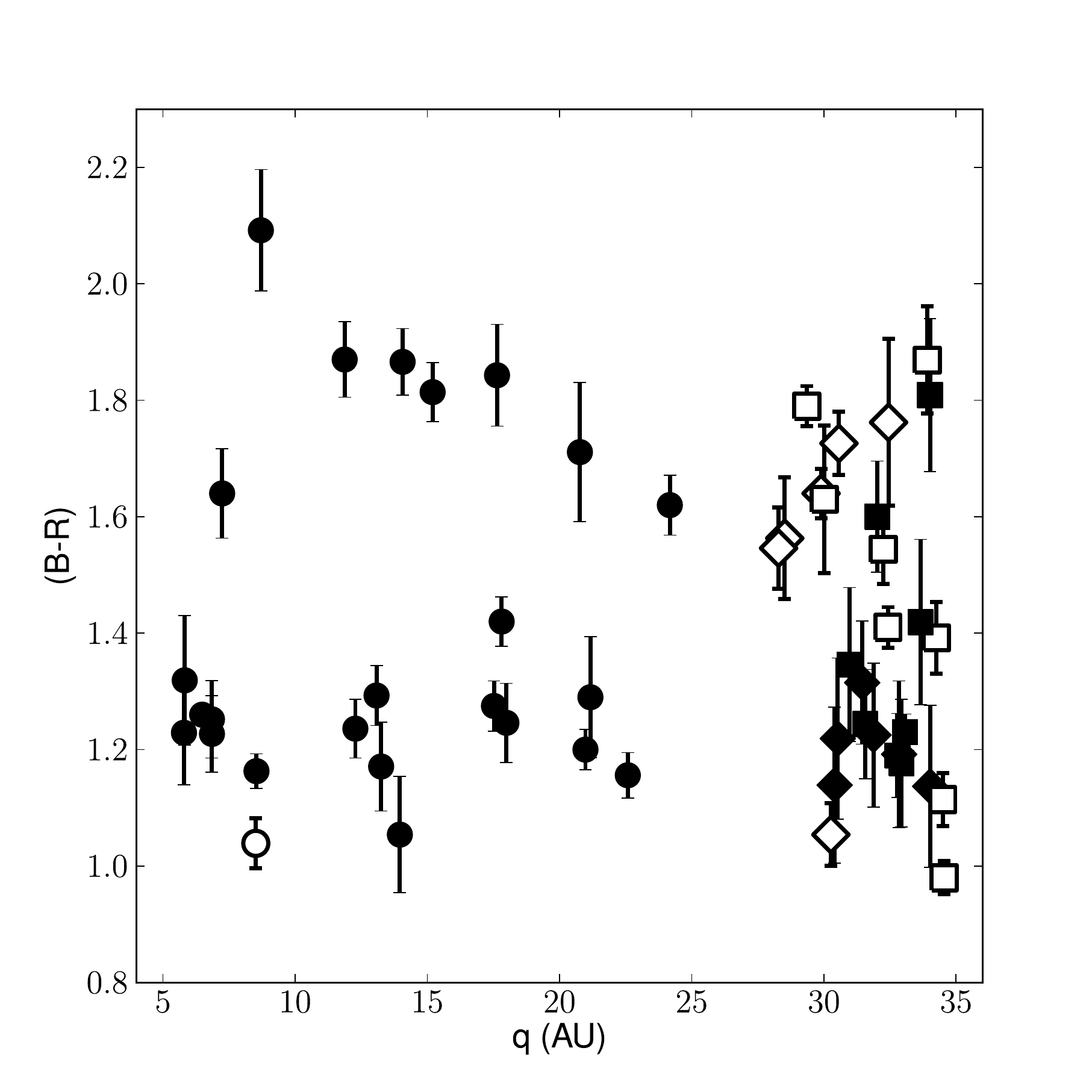} 
   \figcaption{(B-R) colour vs. perihelion of the low-$q$ Centaurs (circles), resonant objects (diamonds) and scattered disk objects (squares) in the MBOSS dataset \citep{Hainaut2002}. Only shown are those objects with measurement errors $\Delta$~(B-R)$<0.15$. Open and closed markers are those objects with  absolute magnitudes larger or smaller than $H_{R}=6.2$ respectively. The bifurcation in (B-R) seen for the Centaurs seems to extend to higher perihelia for small KBOs.   \label{fig:CC_q_MBOSS}}
\end{figure}

We note that the DIP test was sufficient to reveal the bifurcation in the observed ground-based colours of the Centaurs from their optical data alone with a sample size similar to what we present here. This is most likely a result of the use of the (B-R) colour where the bifurcation is most prominent having a width of $\sim 0.3$ magnitudes between the two groups (see Figure~\ref{fig:CC_q_MBOSS}). Thus, further confirmation of our result could be provided with additional observations at $\sim 0.5 \mbox{ $\mu$m}$.

The results of all five separate statistics, the DIP test, the \FOP, and hierarchical, normal mixture model and \textbf{MSDR} clustering, are all in agreement - the colours of the small, excited KBOs are bifurcated. Indeed, the four separate clustering methods produce nearly identical results, with only the normal mixture model results differing from the other three methods, producing the same two main clusters and a third small cluster of outliers. The results of these blind statistical tests support our assertion above, that these objects fall into two compositional classes, the colours of which are described by two branches of a simple two component compositional mixing model. This result solves a long standing issue with the Centaurs, the colours of which appeared different from the other excited populations which are ultimately responsible for replenishing the dynamically short lived Centaurs. We find the simple idea that all small, excited KBOs  fall into one of two compositional classes a pleasing result.

\section{Discussion \label{sec:discussion}}

The primary result from this survey is the demonstration that other than the cold classical objects, the Haumea family members, and those objects with $H<5.6$, the small, excited Kuiper belt objects fall into two separate composition classes, the colours of which are well described by two separate intimate mixtures, both of which share the same neutral component. These results have profound consequences for our understanding of the formation of the Kuiper belt and is indicative of two unique primordial locations from which each class formed.

\subsection{The Colours of the Kuiper belt}

Our observations suggest that the small excited Kuiper belt populations which include the Centaurs, scattered disk, resonant, and hot classical objects fall into two separate compositional classes, the colours of which are determined by the relative composition an object has within it's class. In addition, our observations suggest that little surface evolution occurs as an object evolves from an excited orbit into the Centaur population. This suggests that both classes of objects must have existed within the primordial planetesimal disk before the occurrence of the scattering event which emplaced these objects in the Kuiper belt.

One possible explanation for the primordial existence of two classes of objects is presented in \citet{Brown2011a} who have shown that the methanol sublimation line exists at  $\sim20$ AU, roughly within the primordial disk from which the excited KBOs are believed to have originated. They demonstrated that methanol ice could survive on planetesimal surfaces outside of $\sim 20$ AU while inside this distance, methanol sublimation is rapid enough to deplete all surface methanol before the scattering event occurred. Methanol is an important organic ice, as Solar irradiance can drive methanol chemistry to produce longer chain hydrocarbons.  Surface processing would produce a red irradiated mantel of complex hydrocarbons on roughly Myr timescales, resulting in a red appearance. Objects who spent most of their time inside of 20 AU were unable to retain methanol, resulting in a neutral surface free of a red irradiation mantel. Despite the attractiveness of this model, alternate explanations for the two unique surfaces are possible.

We have found that the blue and red mixing models do not describe the colours of objects with $H_{606}\lesssim5.6$. This roughly corresponds to a diameter of $D\sim 300-400$ km. Our findings are in agreement with albedo observations which suggest that objects larger than $D\sim400$ km have higher albedos than those smaller objects \citep{Stansberry2008}. These results point to a fundamental difference in the surface properties of objects with diameters $D\gtrsim 400$km. These intermediate sized objects also exhibit different surfaces than  the volatile bearing objects with $D\gtrsim 1000$ km. It seems that the surfaces of the intermediate sized objects are dominated by other processes which have little affect at smaller sizes. These could include the onset of differentiation or cryovolcanic processes, both of which would have the effect of covering or removing  non-icy materials from the surface. 

The CCOs exhibit a unique surface with different albedos than either class of excited KBO. As these objects most likely formed nearly in-situ, it appears that the primordial disk from which all KBOs formed generally consisted of three separate classes of small object. The colours of the CCOs  are formally compatible with the red-branch of the mixing model. A simple explanation for this is that both populations formed adjacent to one another in the primordial disk. Due to their larger formation distance however, the classical objects accreted material with a slightly different composition than did the red excited objects. \citet{Brown2011a} suggest that the CCOs may posses a significantly higher fraction of of NH$_3$ compared to the excited populations. While it is not clear what effects variations in NH$_3$ content might have on an object's appearance, it seems reasonable to think that the existence of primordial ices not likely retained by the red excited objects  could result in a different surface.

\subsection{The Nature of the Mixture Components}
Little can be said about the red components of each mixing branch from our observations alone. Other spectroscopic studies however, provide some additional guidance. In modelling observed spectra, the red optical colour of KBOs is usually attributed to irradiated organic material and tholins produced in the laboratory and seen on other icy bodies  \citep[see, for example][]{Cruikshank1991,Brunetto2006,Filacchione2007}. In addition, absorption at $\sim1.5 \mbox{ $\mu$m}$ is often attributed to water-ice. The range of colours compatible with our data are similar to those produced in spectral modelling of these ices \citep{Cruikshank1998,Merlin2010,Barucci2011}. Thus, combinations of irradiated organics and water ice are likely red candidates.

The existence of a common neutral component for both branches implies that this material was prevalent throughout the protoplanetary disc both interior to and exterior to the methanol sublimation line. As discussed above, the best-fit neutral colour is consistent with silicate materials inferred to be present on other KBOs \citep{Merlin2010} including various aqueous altered silicates. Such a result would imply the existence of liquid water in objects in the outer Solar system at some point in the past. Our findings are consistent with other works which suggest the existence of aqueous alteration in KBOs \citep{Alvarez-Candal2008} and comets \citep{Stodolna2010}.  This result is surprising as it is expected that only the largest KBOs would have ever possessed large quantities of liquid water at any stage during their histories \citep{McKinnon2008,Coradini2008}.

One issue still remains with the candidate neutral component, that of albedos. The candidate silicates all have albedos much too high to be compatible with the observed albedos of the small low-$q$ objects. This is consistent with other works which find that to match the low albedos requires the existence of some neutral coloured darkening agent \citep{Cruikshank1998,Barucci2011}. Our results suggest that the relative composition of the darkening agent must not vary significantly compared to the silicate material responsible for the colour of the neutral component. Otherwise, two components (neutral and red) would be insufficient to describe the observed colours of either branch.

Our findings suggest that small excited Kuiper belt objects generally divide into two types. Each type is defined primarily by surface colour, and can be linked to its formation location within the primordial disk and the mechanism responsible for emplacing that object into the Kuiper belt. Previous efforts have been made to classify KBOs based on their surface colours. \citet{Fulchignoni2008} present a recent example in which they used G-mode analysis of available ground-based optical and NIR photometry to identify classes of object. The result was the identification of four separate classes, each with its own unique colour. \citet{Fulchignoni2008} attribute the colours of each of their classes to relative amounts of surface evolution experienced by each class' members. The KBO taxons presented in that and other past efforts are fundamentally different than the types we present here, as the objects in both types represent a range of colours rather than one average value. Our interpretations are also different. The mixing models we consider have testable predictions in regards to the spectral shape of objects amongst each class; as one moves red-ward along a mixing line the spectra of objects at that colour and belonging to that type will vary primarily in a way  consistent with an intimate mixture of two component's spectra. Future detailed spectral observations will be able to test this prediction.

The majority of small excited KBOs are consistent with being drawn from two compositional types. The simple compositional model we present however, has not accounted for a few important effects. The Haumea family demonstrates that collisions can play an important role in modifying the colours of icy bodies. In addition, due to the chaotic nature of the scattering which disrupted the primordial disc, it is possible that a small fraction of Kuiper belt objects originated from different regions than the primary source populations. While these process will modify the surfaces in ways not accounted for by our simple picture, their overall effect is not sufficient to hide the existence of the two unique surface types.

\section{Conclusions \label{sec:conclusions}}
We have presented the first results of the Hubble Wide Field Camera 3 Test of Surfaces in the Outer Solar System. The optical and NIR photometry of this project has revealed that the small KBOs with perihelia $q<35$ AU all exhibit a bifurcated optical colour distribution thought previously to be unique only to the Centaurs. Both the red and blue small KBOs  exhibit highly correlated optical and NIR colours and clearly define two types of KBO surface.  The colours of each surface type are well described by simple two component mixtures models, both of which share a common neutral component but have different red components. The red components are consistent with irradiated organics and tholins seen on other icy bodies. The neutral component is a material that must have pervaded the outer protoplanetary disk, the colours of which are consistent with a range of olivines and aqueously altered silicates.

We have found that the colours of the small $H_{606}>5.6$ objects of all excited dynamical classes are consistent with the colours of our simple mixing model. This suggests that all small excited KBO populations exhibit the same colour distribution, and that the bifurcation observed in the colours of the Centaurs must have existed in the primordial disk before what ever scattering event occurred that populated all excited classes of the Kuiper belt.

The cold classical objects exhibit an additional unique surface. Despite their consistency in colour with red excited objects, their higher albedos and smaller colour range suggest that the cold classical objects and the red excited objects posses different surfaces.

\acknowledgements
We would like to thank Eric Feigelson for his suggestion to use clustering methods in this project. Support for program HST-GO-11644.01-A was provided by NASA through a grant from the Space Telescope Science Institute, which is operated by the Association of Universities for Research in Astronomy, Inc., under NASA contract NAS 5-26555.

\bibliographystyle{apj}
\bibliography{astroelsart}

 \begin{deluxetable}{lcccccccc}
\tabletypesize{\scriptsize}
\tablecaption{Survey Targets and Photometry \label{tab:BIGTABLE}}
\tablehead{\colhead{Designation\tablenotemark{a}} & \colhead{a (AU)} & \colhead{i ($^o$)} & \colhead{e} & \colhead{F606w} & \colhead{F814w} & \colhead{F139m} & \colhead{F153m} & \colhead{H$_{606}$}}
\startdata
\multicolumn{9}{c}{Centaurs}  \\ \hline
Nessus (1993 HA2) & 24.64 & 15.63 & 0.51 & $23.30 \pm 0.04$ & $23.14 \pm 0.04$ & $24.08 \pm 0.05$ & $24.40 \pm 0.05$ & $9.800 \pm 0.04$ \\
Hylonome (1995 DW2) & 25.06 & 4.14 & 0.24 & $23.40 \pm 0.04$ & $23.71 \pm 0.06$ & $24.63 \pm 0.08$ & $24.94 \pm 0.09$ & $10.12 \pm 0.06$ \\
Cyllarus (1998 TF35) & 26.17 & 12.63 & 0.38 & $22.48 \pm 0.02$ & $22.38 \pm 0.02$ & $23.27 \pm 0.02$ & $23.53 \pm 0.03$ & $8.821 \pm 0.02$ \\
29981 (1999 TD10) & 98.15 & 5.95 & 0.87 & $21.98 \pm 0.02$ & $22.21 \pm 0.02$ & $23.36 \pm 0.04$ & $23.56 \pm 0.03$ & $9.032 \pm 0.02$ \\
Elatus (1999 UG5) & 11.76 & 5.24 & 0.38 & $22.43 \pm 0.02$ & $22.46 \pm 0.02$ & $23.64 \pm 0.03$ & $23.92 \pm 0.04$ & $10.92 \pm 0.02$ \\
121725 (1999 XX143) & 17.91 & 6.78 & 0.46 & $23.26 \pm 0.04$ & $23.34 \pm 0.04$ & $24.54 \pm 0.07$ & $24.85 \pm 0.08$ & $9.118 \pm 0.04$ \\
60608 (2000 EE173)\tablenotemark{b} & 48.94 & 5.95 & 0.53 & $21.83 \pm 0.1$ & $22.09 \pm 0.1$ & $23.33 \pm 0.02$ & $23.57 \pm 0.03$ & $8.242 \pm 0.1$ \\
87269 (2000 OO67) & 657.7 & 20.05 & 0.96 & \nodata & \nodata & \nodata & \nodata & \nodata \\
87555 (2000 QB243) & 35.19 & 6.76 & 0.56 & $23.00 \pm 0.03$ & $23.27 \pm 0.04$ & $24.26 \pm 0.06$ & $24.45 \pm 0.06$ & $9.067 \pm 0.04$ \\
63252 (2001 BL41) & 9.729 & 12.45 & 0.29 & $22.95 \pm 0.03$ & $23.22 \pm 0.04$ & $24.08 \pm 0.05$ & $24.33 \pm 0.05$ & $12.05 \pm 0.04$ \\
88269 (2001 KF77) & 26.05 & 4.36 & 0.24 & $23.82 \pm 0.06$ & $23.74 \pm 0.06$ & $24.57 \pm 0.07$ & $25.03 \pm 0.09$ & $10.69 \pm 0.06$ \\
119315 (2001 SQ73) & 17.47 & 17.42 & 0.17 & $22.03 \pm 0.02$ & $22.28 \pm 0.02$ & $23.31 \pm 0.04$ & $23.56 \pm 0.04$ & $9.645 \pm 0.02$ \\
148975 (2001 XA255) & 29.20 & 12.61 & 0.68 & $21.46 \pm 0.02$ & $21.71 \pm 0.02$ & $22.78 \pm 0.02$ & $23.01 \pm 0.03$ & $11.71 \pm 0.02$ \\
Crantor (2002 GO9) & 19.45 & 12.76 & 0.27 & $21.07 \pm 0.02$ & $21.11 \pm 0.01$ & $21.83 \pm 0.01$ & $22.11 \pm 0.01$ & $9.168 \pm 0.01$ \\
2002 QX47 & 25.60 & 7.26 & 0.37 & \nodata & \nodata & $23.42 \pm 0.03$ & $23.69 \pm 0.03$ & \nodata \\
119976 (2002 VR130) & 23.97 & 3.52 & 0.38 & $23.51 \pm 0.04$ & $23.80 \pm 0.06$ & $24.67 \pm 0.08$ & $25.02 \pm 0.09$ & $11.72 \pm 0.06$ \\
127546 (2002 XU93) & 66.44 & 77.96 & 0.68 & $21.72 \pm 0.02$ & $22.04 \pm 0.01$ & $23.08 \pm 0.02$ & $23.29 \pm 0.02$ & $8.497 \pm 0.01$ \\
Ceto (2003 FX128) & 99.85 & 22.32 & 0.82 & $21.85 \pm 0.02$ & $22.02 \pm 0.01$ & $22.96 \pm 0.02$ & $23.26 \pm 0.02$ & $6.888 \pm 0.01$ \\
149560 (2003 QZ91) & 41.86 & 34.76 & 0.47 & $23.13 \pm 0.03$ & $23.38 \pm 0.04$ & $24.28 \pm 0.06$ & $24.52 \pm 0.06$ & $8.800 \pm 0.04$ \\
2004 QQ26 & 22.95 & 21.46 & 0.14 & $23.20 \pm 0.04$ & $23.53 \pm 0.05$ & $24.55 \pm 0.07$ & $24.82 \pm 0.08$ & $10.11 \pm 0.05$ \\
160427 (2005 RL43) & 24.59 & 12.24 & 0.04 & $22.03 \pm 0.02$ & $21.97 \pm 0.01$ & $22.85 \pm 0.02$ & $23.13 \pm 0.02$ & $8.288 \pm 0.01$ \\
2005 RO43 & 28.84 & 35.42 & 0.51 & $21.62 \pm 0.02$ & $21.89 \pm 0.01$ & $22.80 \pm 0.02$ & $23.08 \pm 0.02$ & $7.527 \pm 0.01$ \\
2005 VJ119 & 35.32 & 6.95 & 0.68 & $22.14 \pm 0.02$ & $22.32 \pm 0.03$ & $23.32 \pm 0.04$ & $23.62 \pm 0.04$ & $11.48 \pm 0.03$ \\
2006 QP180 & 38.61 & 4.95 & 0.65 & $22.72 \pm 0.04$ & $22.57 \pm 0.02$ & $23.29 \pm 0.02$ & $23.83 \pm 0.05$ & $10.15 \pm 0.02$ \\
2006 SQ372 & 1079. & 19.45 & 0.97 & $21.94 \pm 0.02$ & $21.95 \pm 0.01$ & $22.98 \pm 0.02$ & $23.21 \pm 0.02$ & $8.132 \pm 0.01$ \\
187661 (2007 JG43) & 24.05 & 33.12 & 0.40 & $21.11 \pm 0.01$ & $21.39 \pm 0.01$ & $22.21 \pm 0.02$ & $22.35 \pm 0.02$ & $9.465 \pm 0.01$ \\
2007 JK43 & 46.35 & 44.89 & 0.49 & $21.22 \pm 0.01$ & $21.41 \pm 0.01$ & $22.37 \pm 0.02$ & $22.62 \pm 0.01$ & $7.442 \pm 0.01$ \\
2007 RH283 & 15.96 & 21.35 & 0.33 & $21.40 \pm 0.01$ & $21.68 \pm 0.01$ & $22.73 \pm 0.02$ & $23.00 \pm 0.02$ & $8.843 \pm 0.01$ \\
\multicolumn{9}{c}{Scattered}  \\ \hline
26308 (1998 SM165) & 47.99 & 13.47 & 0.37 & $22.23 \pm 0.03$ & $22.13 \pm 0.02$ & $22.91 \pm 0.02$ & $23.18 \pm 0.03$ & $6.466 \pm 0.02$ \\
69986 (1998 WW24) & 39.49 & 13.93 & 0.22 & $23.88 \pm 0.08$ & $24.01 \pm 0.08$ & $24.86 \pm 0.10$ & $25.20 \pm 0.11$ & $8.695 \pm 0.08$ \\
40314 (1999 KR16) & 48.83 & 24.84 & 0.30 & $21.61 \pm 0.02$ & $21.48 \pm 0.01$ & $22.36 \pm 0.01$ & $22.68 \pm 0.01$ & $6.107 \pm 0.01$ \\
1999 RJ215 & 59.91 & 19.68 & 0.42 & $23.06 \pm 0.03$ & $23.22 \pm 0.04$ & $24.13 \pm 0.05$ & $24.44 \pm 0.06$ & $7.706 \pm 0.04$ \\
86177 (1999 RY215) & 45.59 & 22.13 & 0.24 & $22.82 \pm 0.03$ & $23.17 \pm 0.04$ & $24.26 \pm 0.06$ & $24.43 \pm 0.05$ & $7.313 \pm 0.04$ \\
91554 (1999 RZ215) & 104.0 & 25.46 & 0.70 & $23.13 \pm 0.03$ & $23.35 \pm 0.04$ & $24.26 \pm 0.06$ & $24.58 \pm 0.06$ & $7.972 \pm 0.04$ \\
2000 AF255 & 48.43 & 30.87 & 0.24 & $23.38 \pm 0.04$ & $23.17 \pm 0.04$ & $24.05 \pm 0.05$ & $24.40 \pm 0.05$ & $6.305 \pm 0.04$ \\
2000 CQ105 & 56.74 & 19.72 & 0.39 & $23.26 \pm 0.04$ & $23.61 \pm 0.05$ & $24.66 \pm 0.08$ & $24.92 \pm 0.09$ & $6.548 \pm 0.05$ \\
130391 (2000 JG81) & 47.52 & 23.46 & 0.28 & $23.72 \pm 0.05$ & $23.90 \pm 0.07$ & $24.79 \pm 0.09$ & $25.09 \pm 0.10$ & $8.301 \pm 0.07$ \\
2000 QL251 & 48.12 & 3.67 & 0.22 & \nodata & \nodata & \nodata & \nodata & \nodata \\
2000 YB2 & 38.76 & 3.82 & 0.03 & $23.05 \pm 0.03$ & $23.23 \pm 0.06$ & $24.14 \pm 0.05$ & $24.48 \pm 0.06$ & $7.198 \pm 0.06$ \\
2000 YH2 & 39.22 & 12.90 & 0.30 & $22.63 \pm 0.04$ & $22.95 \pm 0.03$ & $24.21 \pm 0.08$ & $24.42 \pm 0.05$ & $8.273 \pm 0.03$ \\
82158 (2001 FP185) & 214.0 & 30.83 & 0.83 & \nodata & \nodata & $23.03 \pm 0.02$ & $23.33 \pm 0.02$ & \nodata \\
82155 (2001 FZ173) & 84.97 & 12.73 & 0.61 & \nodata & \nodata & $22.86 \pm 0.02$ & $23.15 \pm 0.02$ & \nodata \\
2001 QR322 & 30.36 & 1.32 & 0.03 & $23.21 \pm 0.04$ & $23.51 \pm 0.05$ & $24.51 \pm 0.07$ & $24.79 \pm 0.08$ & $8.510 \pm 0.05$ \\
2001 QX322 & 58.49 & 28.53 & 0.39 & $22.97 \pm 0.03$ & $23.16 \pm 0.04$ & $24.15 \pm 0.05$ & $24.46 \pm 0.06$ & $6.842 \pm 0.04$ \\
2001 UP18 & 47.94 & 1.17 & 0.07 & \nodata & \nodata & $24.56 \pm 0.07$ & $24.82 \pm 0.08$ & \nodata \\
126155 (2001 YJ140) & 39.37 & 5.98 & 0.29 & $22.17 \pm 0.02$ & $22.44 \pm 0.02$ & $23.38 \pm 0.03$ & $23.57 \pm 0.03$ & $7.764 \pm 0.02$ \\
119979 (2002 WC19) & 47.80 & 9.19 & 0.25 & $21.14 \pm 0.01$ & $21.19 \pm 0.01$ & $22.12 \pm 0.01$ & $22.38 \pm 0.01$ & $4.879 \pm 0.01$ \\
2003 FE128 & 47.92 & 3.38 & 0.25 & $22.28 \pm 0.02$ & $22.34 \pm 0.02$ & $23.41 \pm 0.03$ & $23.67 \pm 0.03$ & $6.781 \pm 0.02$ \\
2003 FF128 & 39.54 & 1.91 & 0.21 & $22.35 \pm 0.02$ & $22.34 \pm 0.02$ & $23.35 \pm 0.03$ & $23.69 \pm 0.04$ & $7.319 \pm 0.02$ \\
2003 FJ127 & 44.02 & 22.93 & 0.22 & \nodata & \nodata & \nodata & \nodata & \nodata \\
2003 QA92 & 38.45 & 3.42 & 0.06 & $22.57 \pm 0.02$ & $22.62 \pm 0.02$ & $23.62 \pm 0.03$ & $23.84 \pm 0.03$ & $6.880 \pm 0.02$ \\
2003 QX91 & 43.99 & 27.62 & 0.25 & $24.25 \pm 0.08$ & $24.49 \pm 0.12$ & $25.26 \pm 0.14$ & $25.66 \pm 0.17$ & $9.007 \pm 0.12$ \\
2003 UY413 & 67.31 & 20.58 & 0.57 & \nodata & \nodata & \nodata & \nodata & \nodata \\
Sedna (2003 VB12) & 510.2 & 11.92 & 0.85 & \nodata & \nodata & $21.99 \pm 0.01$ & $22.31 \pm 0.01$ & \nodata \\
2003 WU172 & 39.16 & 4.14 & 0.25 & $21.36 \pm 0.01$ & $21.46 \pm 0.01$ & $22.47 \pm 0.01$ & $22.74 \pm 0.02$ & $6.656 \pm 0.01$ \\
120216 (2004 EW95) & 39.36 & 29.30 & 0.31 & $21.18 \pm 0.01$ & $21.56 \pm 0.01$ & $22.68 \pm 0.01$ & $22.98 \pm 0.02$ & $6.796 \pm 0.01$ \\
2004 PA108 & 54.57 & 1.14 & 0.46 & \nodata & \nodata & \nodata & \nodata & \nodata \\
2004 PA112 & 39.14 & 32.86 & 0.11 & $23.22 \pm 0.04$ & $23.55 \pm 0.05$ & $24.42 \pm 0.06$ & $24.63 \pm 0.07$ & $7.784 \pm 0.05$ \\
2004 VN112 & 341.8 & 25.54 & 0.86 & \nodata & \nodata & $24.74 \pm 0.09$ & $24.95 \pm 0.09$ & \nodata \\
2004 XR190 & 57.47 & 46.66 & 0.10 & $22.14 \pm 0.02$ & $22.42 \pm 0.02$ & $23.46 \pm 0.03$ & $23.74 \pm 0.03$ & $4.545 \pm 0.02$ \\
2005 EB299 & 51.76 & 0.71 & 0.50 & \nodata & \nodata & \nodata & \nodata & \nodata \\
2005 GE187 & 39.44 & 18.26 & 0.32 & $22.56 \pm 0.02$ & $22.55 \pm 0.02$ & $23.40 \pm 0.03$ & $23.62 \pm 0.03$ & $7.850 \pm 0.02$ \\
2005 GF187 & 39.50 & 3.90 & 0.25 & $23.07 \pm 0.03$ & $23.43 \pm 0.05$ & $24.49 \pm 0.07$ & $24.77 \pm 0.07$ & $8.284 \pm 0.05$ \\
2005 PU21 & 179.4 & 6.16 & 0.83 & \nodata & \nodata & $24.15 \pm 0.05$ & $24.39 \pm 0.05$ & \nodata \\
2005 RS43 & 48.11 & 9.99 & 0.20 & $21.72 \pm 0.02$ & $22.00 \pm 0.01$ & $23.11 \pm 0.02$ & $23.37 \pm 0.02$ & $5.464 \pm 0.01$ \\
145474 (2005 SA278) & 92.90 & 16.25 & 0.64 & $22.73 \pm 0.03$ & $23.03 \pm 0.03$ & $24.21 \pm 0.05$ & $24.48 \pm 0.06$ & $6.724 \pm 0.03$ \\
145480 (2005 TB190) & 76.58 & 26.43 & 0.39 & $21.32 \pm 0.01$ & $21.46 \pm 0.01$ & $22.59 \pm 0.01$ & $22.90 \pm 0.01$ & $4.703 \pm 0.01$ \\
2005 TV189 & 39.42 & 34.39 & 0.18 & $22.95 \pm 0.03$ & $23.21 \pm 0.04$ & $24.08 \pm 0.05$ & $24.31 \pm 0.05$ & $7.967 \pm 0.04$ \\
2006 QH181 & 67.51 & 19.22 & 0.43 & $23.76 \pm 0.05$ & $23.80 \pm 0.06$ & $24.46 \pm 0.07$ & $24.63 \pm 0.07$ & $4.640 \pm 0.06$ \\
225088 (2007 OR10) & 67.34 & 30.67 & 0.49 & $21.67 \pm 0.02$ & $21.36 \pm 0.01$ & $22.06 \pm 0.01$ & $22.46 \pm 0.01$ & $2.335 \pm 0.01$ \\
2007 TA418 & 72.92 & 21.96 & 0.50 & $23.26 \pm 0.04$ & $23.46 \pm 0.05$ & $24.41 \pm 0.06$ & $24.73 \pm 0.07$ & $7.622 \pm 0.05$ \\
\multicolumn{9}{c}{Resonant\tablenotemark{c}}  \\ \hline
118228 (1996 TQ66) & 39.51 & 14.64 & 0.12 & \nodata & \nodata & \nodata & \nodata & \nodata \\
137295 (1999 RB216) & 47.87 & 12.66 & 0.29 & $22.68 \pm 0.03$ & $22.82 \pm 0.03$ & $23.91 \pm 0.04$ & $24.15 \pm 0.04$ & $7.456 \pm 0.03$ \\
60620 (2000 FD8) & 43.68 & 19.54 & 0.22 & $22.86 \pm 0.03$ & $22.83 \pm 0.03$ & $23.54 \pm 0.03$ & $23.91 \pm 0.03$ & $7.047 \pm 0.03$ \\
2000 FV53 & 39.18 & 17.35 & 0.16 & $23.25 \pm 0.04$ & $23.69 \pm 0.06$ & $24.69 \pm 0.08$ & $24.85 \pm 0.08$ & $8.020 \pm 0.06$ \\
131318 (2001 FL194) & 39.22 & 13.72 & 0.17 & $23.32 \pm 0.04$ & $23.60 \pm 0.05$ & $24.65 \pm 0.08$ & $24.80 \pm 0.08$ & $8.152 \pm 0.05$ \\
2001 FQ185 & 47.57 & 3.24 & 0.22 & $23.25 \pm 0.04$ & $23.12 \pm 0.03$ & $24.11 \pm 0.05$ & $24.33 \pm 0.05$ & $7.613 \pm 0.03$ \\
139775 (2001 QG298) & 39.61 & 6.48 & 0.19 & $22.54 \pm 0.02$ & $22.56 \pm 0.02$ & $23.59 \pm 0.05$ & $23.88 \pm 0.05$ & $7.590 \pm 0.02$ \\
2002 GY32 & 39.45 & 1.80 & 0.08 & $23.30 \pm 0.04$ & $23.38 \pm 0.04$ & $24.38 \pm 0.06$ & $24.60 \pm 0.06$ & $7.757 \pm 0.04$ \\
2004 EH96 & 39.30 & 3.13 & 0.27 & $23.19 \pm 0.03$ & $23.11 \pm 0.03$ & $24.17 \pm 0.07$ & $24.44 \pm 0.06$ & $8.555 \pm 0.03$ \\
2004 TV357 & 47.81 & 9.76 & 0.27 & $22.55 \pm 0.02$ & $23.00 \pm 0.03$ & $24.11 \pm 0.05$ & $24.39 \pm 0.05$ & $7.031 \pm 0.03$ \\
2005 CA79 & 47.57 & 11.66 & 0.22 & $21.29 \pm 0.01$ & $21.45 \pm 0.01$ & $22.46 \pm 0.01$ & $22.72 \pm 0.01$ & $5.606 \pm 0.01$ \\
2005 EZ296 & 39.35 & 1.77 & 0.14 & $22.89 \pm 0.03$ & $22.89 \pm 0.03$ & $23.91 \pm 0.04$ & $24.23 \pm 0.05$ & $7.456 \pm 0.03$ \\
2005 GB187 & 39.43 & 14.69 & 0.23 & $22.25 \pm 0.02$ & $22.52 \pm 0.03$ & \nodata & \nodata & $7.415 \pm 0.03$ \\
\multicolumn{9}{c}{Hot Classical}  \\ \hline
1999 CL119 & 46.71 & 23.34 & 0.00 & $23.08 \pm 0.03$ & $23.18 \pm 0.04$ & $24.40 \pm 0.06$ & $24.66 \pm 0.07$ & $6.416 \pm 0.04$ \\
2000 OH67 & 44.44 & 5.62 & 0.02 & $24.12 \pm 0.07$ & $24.17 \pm 0.09$ & $25.06 \pm 0.16$ & $25.39 \pm 0.19$ & $7.751 \pm 0.09$ \\
150642 (2001 CZ31) & 45.00 & 10.23 & 0.11 & $22.20 \pm 0.02$ & $22.58 \pm 0.02$ & $23.60 \pm 0.03$ & $23.84 \pm 0.03$ & $6.122 \pm 0.02$ \\
2001 FO185 & 46.39 & 10.65 & 0.11 & \nodata & \nodata & \nodata & \nodata & \nodata \\
2001 HY65 & 43.06 & 17.16 & 0.11 & $22.64 \pm 0.02$ & $22.73 \pm 0.02$ & $23.73 \pm 0.04$ & $24.03 \pm 0.04$ & $6.689 \pm 0.02$ \\
2001 KA77 & 47.56 & 11.91 & 0.09 & $22.61 \pm 0.02$ & $22.50 \pm 0.02$ & $23.39 \pm 0.04$ & $23.70 \pm 0.04$ & $5.779 \pm 0.02$ \\
2001 PK47 & 39.93 & 8.73 & 0.06 & $23.48 \pm 0.04$ & $23.63 \pm 0.05$ & $24.65 \pm 0.08$ & $24.75 \pm 0.07$ & $7.748 \pm 0.05$ \\
2001 QC298 & 46.56 & 30.50 & 0.12 & $22.68 \pm 0.03$ & $23.11 \pm 0.05$ & $24.14 \pm 0.05$ & $24.41 \pm 0.05$ & $6.639 \pm 0.05$ \\
2001 QR297 & 44.69 & 5.13 & 0.03 & $23.47 \pm 0.04$ & $23.42 \pm 0.04$ & $24.41 \pm 0.06$ & $24.58 \pm 0.09$ & $7.068 \pm 0.04$ \\
2002 PD155 & 43.49 & 5.75 & 0.01 & $23.97 \pm 0.06$ & $24.22 \pm 0.13$ & $24.93 \pm 0.10$ & $25.30 \pm 0.12$ & $7.687 \pm 0.13$ \\
2003 LD9 & 47.36 & 6.97 & 0.17 & $23.35 \pm 0.04$ & $23.30 \pm 0.04$ & $24.38 \pm 0.06$ & $24.62 \pm 0.07$ & $7.198 \pm 0.04$ \\
2003 QF113 & 44.10 & 4.45 & 0.03 & $23.61 \pm 0.05$ & $23.59 \pm 0.05$ & $24.73 \pm 0.09$ & $24.88 \pm 0.08$ & $7.348 \pm 0.05$ \\
2003 SQ317 & 42.87 & 28.51 & 0.08 & $22.70 \pm 0.03$ & $23.22 \pm 0.06$ & $24.64 \pm 0.08$ & $26.01 \pm 0.23$ & $6.818 \pm 0.06$ \\
2003 UZ117 & 44.29 & 27.39 & 0.13 & $21.39 \pm 0.01$ & $21.87 \pm 0.01$ & $23.39 \pm 0.03$ & $24.62 \pm 0.06$ & $5.462 \pm 0.01$ \\
90568 (2004 GV9) & 41.90 & 21.97 & 0.07 & $20.25 \pm 0.01$ & $20.391 \pm 0.009$ & $21.45 \pm 0.01$ & $21.70 \pm 0.01$ & $4.369 \pm 0.00$ \\
145452 (2005 RN43) & 41.77 & 19.24 & 0.02 & $20.16 \pm 0.01$ & $20.298 \pm 0.009$ & $21.40 \pm 0.01$ & $21.65 \pm 0.01$ & $4.044 \pm 0.00$ \\
\multicolumn{9}{c}{Cold Classical}  \\ \hline
1998 WY24 & 43.14 & 1.91 & 0.03 & $23.57 \pm 0.05$ & $23.64 \pm 0.05$ & $24.61 \pm 0.08$ & $24.88 \pm 0.08$ & $7.394 \pm 0.05$ \\
2000 CE105 & 43.72 & 0.54 & 0.05 & $23.90 \pm 0.06$ & $23.81 \pm 0.06$ & $24.80 \pm 0.09$ & $24.98 \pm 0.09$ & $7.762 \pm 0.06$ \\
123509 (2000 WK183) & 44.37 & 1.96 & 0.04 & $23.11 \pm 0.03$ & $23.19 \pm 0.04$ & $24.22 \pm 0.05$ & $24.46 \pm 0.06$ & $6.814 \pm 0.04$ \\
2000 WT169 & 44.87 & 1.74 & 0.01 & $22.80 \pm 0.03$ & $22.87 \pm 0.03$ & $23.83 \pm 0.04$ & $24.13 \pm 0.04$ & $6.303 \pm 0.03$ \\
2000 WW12 & 44.74 & 2.49 & 0.12 & \nodata & \nodata & \nodata & \nodata & \nodata \\
88268 (2001 KK76) & 42.60 & 1.88 & 0.02 & $23.33 \pm 0.04$ & $23.26 \pm 0.04$ & $24.14 \pm 0.05$ & $24.29 \pm 0.05$ & $7.093 \pm 0.04$ \\
160147 (2001 KN76) & 43.71 & 2.64 & 0.08 & $23.32 \pm 0.04$ & $23.38 \pm 0.04$ & $24.47 \pm 0.07$ & $24.70 \pm 0.07$ & $7.310 \pm 0.04$ \\
2001 OQ108 & 45.52 & 2.32 & 0.00 & $23.79 \pm 0.05$ & $23.72 \pm 0.06$ & $24.73 \pm 0.09$ & $24.97 \pm 0.09$ & $7.231 \pm 0.06$ \\
2001 QS322 & 44.28 & 0.24 & 0.04 & $23.65 \pm 0.05$ & $23.67 \pm 0.06$ & $24.71 \pm 0.08$ & $24.87 \pm 0.08$ & $7.413 \pm 0.06$ \\
Teharonhiawako (2001 QT297) & 44.29 & 2.57 & 0.02 & $22.83 \pm 0.03$ & $22.91 \pm 0.04$ & $23.97 \pm 0.04$ & $24.16 \pm 0.04$ & $6.302 \pm 0.04$ \\
2001 QX297 & 44.42 & 0.90 & 0.03 & $23.42 \pm 0.04$ & $23.35 \pm 0.04$ & $24.37 \pm 0.06$ & $24.67 \pm 0.07$ & $7.060 \pm 0.04$ \\
2001 RW143 & 43.21 & 2.96 & 0.03 & $23.90 \pm 0.06$ & $24.01 \pm 0.08$ & \nodata & \nodata & $7.711 \pm 0.08$ \\
2001 RZ143 & 44.23 & 2.11 & 0.06 & $22.99 \pm 0.03$ & $23.06 \pm 0.03$ & $24.05 \pm 0.05$ & $24.33 \pm 0.05$ & $6.883 \pm 0.03$ \\
2002 CU154 & 43.55 & 3.35 & 0.05 & $23.64 \pm 0.05$ & $23.65 \pm 0.06$ & $24.76 \pm 0.09$ & $25.18 \pm 0.11$ & $7.546 \pm 0.06$ \\
2002 FW36 & 42.74 & 2.35 & 0.02 & \nodata & \nodata & $24.84 \pm 0.09$ & $25.15 \pm 0.11$ & \nodata \\
2002 PV170 & 42.93 & 1.27 & 0.01 & $22.80 \pm 0.03$ & $22.80 \pm 0.03$ & $23.83 \pm 0.04$ & $24.09 \pm 0.04$ & $6.530 \pm 0.03$ \\
2003 GH55 & 44.12 & 1.10 & 0.07 & $22.63 \pm 0.02$ & $22.65 \pm 0.02$ & $23.53 \pm 0.03$ & $23.79 \pm 0.03$ & $6.529 \pm 0.02$ \\
2003 HG57 & 43.55 & 2.10 & 0.03 & $22.70 \pm 0.04$ & $22.75 \pm 0.03$ & $23.94 \pm 0.06$ & $23.99 \pm 0.05$ & $6.368 \pm 0.03$ \\
2003 QZ111 & 43.27 & 2.65 & 0.06 & \nodata & \nodata & \nodata & \nodata & \nodata \\
2006 HW122 & 45.48 & 1.53 & 0.06 & $23.99 \pm 0.06$ & $24.04 \pm 0.08$ & $24.97 \pm 0.11$ & $25.46 \pm 0.14$ & $7.558 \pm 0.08$ \\
\enddata
\tablenotetext{1}{All magnitudes are presented in the STMag system.}
\tablenotetext{a}{MPC designations or names and asteroid numbers (where available). H$_{606}$ is the absolute magnitude of the object determined from its observed magnitude in the F606w filter and its distance at time of observation.}
\tablenotetext{b}{F606w and F814w photometry determined from observations presented by \citet{Benecchi2011} (see Section~\ref{sec:observing}).}
\tablenotetext{c}{Resonant classifications generously provided by Brett Gladman \citep{Gladman2008}.}
\end{deluxetable}

 \begin{deluxetable}{lccc}
\tabletypesize{\scriptsize}
\tablecaption{Observed Colours\label{tab:BIGTABLE2}}
\tablehead{\colhead{Designation} &  \colhead{F606w-F814w} & \colhead{F814w-F139m} & \colhead{F139m-F153m}}
\startdata
\multicolumn{4}{c}{Centaurs}  \\ \hline
Nessus (1993 HA2) & $0.156 \pm 0.05$ & $-0.94 \pm 0.06$ & $-0.31 \pm 0.07$ \\
Hylonome (1995 DW2) & $-0.31 \pm 0.07$ & $-0.91 \pm 0.10$ & $-0.31 \pm 0.12$ \\
Cyllarus (1998 TF35) & $0.095 \pm 0.03$ & $-0.88 \pm 0.03$ & $-0.25 \pm 0.04$ \\
29981 (1999 TD10) & $-0.23 \pm 0.03$ & $-1.14 \pm 0.04$ & $-0.20 \pm 0.05$ \\
Elatus (1999 UG5) & $-0.02 \pm 0.03$ & $-1.18 \pm 0.04$ & $-0.27 \pm 0.05$ \\
121725 (1999 XX143) & $-0.07 \pm 0.06$ & $-1.20 \pm 0.08$ & $-0.30 \pm 0.11$ \\
60608 (2000 EE173)\tablenotemark{b} & $-0.26 \pm 0.14$ & $-1.23 \pm 0.10$ & $-0.23 \pm 0.04$ \\
87269 (2000 OO67) & \nodata & \nodata & \nodata \\
87555 (2000 QB243) & $-0.26 \pm 0.05$ & $-0.99 \pm 0.07$ & $-0.18 \pm 0.08$ \\
63252 (2001 BL41) & $-0.27 \pm 0.05$ & $-0.85 \pm 0.06$ & $-0.25 \pm 0.07$ \\
88269 (2001 KF77) & $0.085 \pm 0.08$ & $-0.83 \pm 0.10$ & $-0.46 \pm 0.12$ \\
119315 (2001 SQ73) & $-0.25 \pm 0.03$ & $-1.02 \pm 0.04$ & $-0.25 \pm 0.06$ \\
148975 (2001 XA255) & $-0.24 \pm 0.03$ & $-1.07 \pm 0.03$ & $-0.22 \pm 0.04$ \\
Crantor (2002 GO9) & $-0.04 \pm 0.03$ & $-0.71 \pm 0.02$ & $-0.28 \pm 0.01$ \\
2002 QX47 & \nodata & \nodata & $-0.26 \pm 0.04$ \\
119976 (2002 VR130) & $-0.28 \pm 0.08$ & $-0.87 \pm 0.10$ & $-0.34 \pm 0.13$ \\
127546 (2002 XU93) & $-0.32 \pm 0.02$ & $-1.03 \pm 0.03$ & $-0.20 \pm 0.03$ \\
Ceto (2003 FX128) & $-0.17 \pm 0.02$ & $-0.94 \pm 0.02$ & $-0.30 \pm 0.03$ \\
149560 (2003 QZ91) & $-0.24 \pm 0.06$ & $-0.90 \pm 0.07$ & $-0.23 \pm 0.08$ \\
2004 QQ26 & $-0.33 \pm 0.06$ & $-1.01 \pm 0.09$ & $-0.27 \pm 0.11$ \\
160427 (2005 RL43) & $0.065 \pm 0.02$ & $-0.88 \pm 0.02$ & $-0.28 \pm 0.03$ \\
2005 RO43 & $-0.27 \pm 0.02$ & $-0.90 \pm 0.02$ & $-0.27 \pm 0.03$ \\
2005 VJ119 & $-0.18 \pm 0.04$ & $-0.99 \pm 0.05$ & $-0.30 \pm 0.06$ \\
2006 QP180 & $0.146 \pm 0.05$ & $-0.71 \pm 0.03$ & $-0.53 \pm 0.06$ \\
2006 SQ372 & $-0.00 \pm 0.02$ & $-1.03 \pm 0.02$ & $-0.22 \pm 0.03$ \\
187661 (2007 JG43) & $-0.27 \pm 0.02$ & $-0.82 \pm 0.02$ & $-0.13 \pm 0.03$ \\
2007 JK43 & $-0.19 \pm 0.02$ & $-0.96 \pm 0.02$ & $-0.24 \pm 0.02$ \\
2007 RH283 & $-0.28 \pm 0.02$ & $-1.04 \pm 0.02$ & $-0.27 \pm 0.02$ \\
\multicolumn{4}{c}{Scattered}  \\ \hline
26308 (1998 SM165) & $0.104 \pm 0.04$ & $-0.78 \pm 0.03$ & $-0.27 \pm 0.04$ \\
69986 (1998 WW24) & $-0.13 \pm 0.11$ & $-0.84 \pm 0.12$ & $-0.34 \pm 0.15$ \\
40314 (1999 KR16) & $0.125 \pm 0.02$ & $-0.87 \pm 0.02$ & $-0.32 \pm 0.02$ \\
1999 RJ215 & $-0.16 \pm 0.05$ & $-0.91 \pm 0.06$ & $-0.30 \pm 0.08$ \\
86177 (1999 RY215) & $-0.34 \pm 0.05$ & $-1.08 \pm 0.07$ & $-0.16 \pm 0.08$ \\
91554 (1999 RZ215) & $-0.21 \pm 0.06$ & $-0.90 \pm 0.07$ & $-0.32 \pm 0.09$ \\
2000 AF255 & $0.211 \pm 0.06$ & $-0.88 \pm 0.06$ & $-0.34 \pm 0.07$ \\
2000 CQ105 & $-0.34 \pm 0.07$ & $-1.04 \pm 0.10$ & $-0.26 \pm 0.12$ \\
130391 (2000 JG81) & $-0.18 \pm 0.09$ & $-0.88 \pm 0.12$ & $-0.29 \pm 0.14$ \\
2000 QL251 & \nodata & \nodata & \nodata \\
2000 YB2 & $-0.18 \pm 0.07$ & $-0.90 \pm 0.08$ & $-0.33 \pm 0.08$ \\
2000 YH2 & $-0.32 \pm 0.05$ & $-1.25 \pm 0.08$ & $-0.21 \pm 0.10$ \\
82158 (2001 FP185) & \nodata & \nodata & $-0.29 \pm 0.03$ \\
82155 (2001 FZ173) & \nodata & \nodata & $-0.29 \pm 0.03$ \\
2001 QR322 & $-0.29 \pm 0.06$ & $-0.99 \pm 0.09$ & $-0.28 \pm 0.11$ \\
2001 QX322 & $-0.18 \pm 0.05$ & $-0.99 \pm 0.06$ & $-0.30 \pm 0.08$ \\
2001 UP18 & \nodata & \nodata & $-0.26 \pm 0.11$ \\
126155 (2001 YJ140) & $-0.26 \pm 0.03$ & $-0.93 \pm 0.03$ & $-0.19 \pm 0.04$ \\
119979 (2002 WC19) & $-0.04 \pm 0.02$ & $-0.93 \pm 0.01$ & $-0.25 \pm 0.02$ \\
2003 FE128 & $-0.05 \pm 0.03$ & $-1.06 \pm 0.03$ & $-0.26 \pm 0.04$ \\
2003 FF128 & $0.007 \pm 0.03$ & $-1.01 \pm 0.03$ & $-0.33 \pm 0.05$ \\
2003 FJ127 & \nodata & \nodata & \nodata \\
2003 QA92 & $-0.05 \pm 0.03$ & $-0.99 \pm 0.04$ & $-0.22 \pm 0.05$ \\
2003 QX91 & $-0.24 \pm 0.14$ & $-0.76 \pm 0.18$ & $-0.40 \pm 0.22$ \\
2003 UY413 & \nodata & \nodata & \nodata \\
Sedna (2003 VB12) & \nodata & \nodata & $-0.31 \pm 0.02$ \\
2003 WU172 & $-0.09 \pm 0.02$ & $-1.00 \pm 0.02$ & $-0.27 \pm 0.03$ \\
120216 (2004 EW95) & $-0.38 \pm 0.02$ & $-1.11 \pm 0.02$ & $-0.29 \pm 0.02$ \\
2004 PA108 & \nodata & \nodata & \nodata \\
2004 PA112 & $-0.32 \pm 0.06$ & $-0.87 \pm 0.08$ & $-0.20 \pm 0.09$ \\
2004 VN112 & \nodata & \nodata & $-0.21 \pm 0.12$ \\
2004 XR190 & $-0.27 \pm 0.03$ & $-1.04 \pm 0.04$ & $-0.28 \pm 0.04$ \\
2005 EB299 & \nodata & \nodata & \nodata \\
2005 GE187 & $0.014 \pm 0.03$ & $-0.85 \pm 0.04$ & $-0.21 \pm 0.04$ \\
2005 GF187 & $-0.36 \pm 0.06$ & $-1.05 \pm 0.08$ & $-0.27 \pm 0.10$ \\
2005 PU21 & \nodata & \nodata & $-0.23 \pm 0.08$ \\
2005 RS43 & $-0.27 \pm 0.02$ & $-1.10 \pm 0.03$ & $-0.26 \pm 0.03$ \\
145474 (2005 SA278) & $-0.29 \pm 0.04$ & $-1.17 \pm 0.06$ & $-0.26 \pm 0.08$ \\
145480 (2005 TB190) & $-0.13 \pm 0.02$ & $-1.13 \pm 0.02$ & $-0.31 \pm 0.02$ \\
2005 TV189 & $-0.25 \pm 0.05$ & $-0.87 \pm 0.06$ & $-0.22 \pm 0.07$ \\
2006 QH181 & $-0.03 \pm 0.08$ & $-0.66 \pm 0.09$ & $-0.16 \pm 0.10$ \\
225088 (2007 OR10) & $0.301 \pm 0.02$ & $-0.69 \pm 0.01$ & $-0.40 \pm 0.02$ \\
2007 TA418 & $-0.20 \pm 0.06$ & $-0.95 \pm 0.08$ & $-0.32 \pm 0.10$ \\
\multicolumn{4}{c}{Resonant}  \\ \hline
118228 (1996 TQ66) & \nodata & \nodata & \nodata \\
137295 (1999 RB216) & $-0.14 \pm 0.04$ & $-1.09 \pm 0.05$ & $-0.24 \pm 0.06$ \\
60620 (2000 FD8) & $0.039 \pm 0.04$ & $-0.71 \pm 0.04$ & $-0.37 \pm 0.05$ \\
2000 FV53 & $-0.43 \pm 0.07$ & $-0.99 \pm 0.10$ & $-0.15 \pm 0.12$ \\
131318 (2001 FL194) & $-0.28 \pm 0.07$ & $-1.04 \pm 0.10$ & $-0.15 \pm 0.11$ \\
2001 FQ185 & $0.133 \pm 0.05$ & $-0.98 \pm 0.06$ & $-0.22 \pm 0.07$ \\
139775 (2001 QG298) & $-0.02 \pm 0.03$ & $-1.02 \pm 0.05$ & $-0.28 \pm 0.07$ \\
2002 GY32 & $-0.08 \pm 0.06$ & $-0.99 \pm 0.08$ & $-0.22 \pm 0.09$ \\
2004 EH96 & $0.078 \pm 0.05$ & $-1.05 \pm 0.08$ & $-0.27 \pm 0.10$ \\
2004 TV357 & $-0.45 \pm 0.04$ & $-1.10 \pm 0.06$ & $-0.28 \pm 0.07$ \\
2005 CA79 & $-0.15 \pm 0.02$ & $-1.01 \pm 0.02$ & $-0.26 \pm 0.02$ \\
2005 EZ296 & $0.005 \pm 0.04$ & $-1.02 \pm 0.05$ & $-0.31 \pm 0.06$ \\
2005 GB187 & $-0.27 \pm 0.04$ & \nodata & \nodata \\
\multicolumn{4}{c}{Hot Classical}  \\ \hline
1999 CL119 & $-0.10 \pm 0.05$ & $-1.21 \pm 0.08$ & $-0.25 \pm 0.09$ \\
2000 OH67 & $-0.04 \pm 0.11$ & $-0.89 \pm 0.19$ & $-0.33 \pm 0.25$ \\
150642 (2001 CZ31) & $-0.37 \pm 0.03$ & $-1.01 \pm 0.04$ & $-0.24 \pm 0.05$ \\
2001 FO185 & \nodata & \nodata & \nodata \\
2001 HY65 & $-0.08 \pm 0.04$ & $-0.99 \pm 0.04$ & $-0.30 \pm 0.05$ \\
2001 KA77 & $0.113 \pm 0.03$ & $-0.89 \pm 0.05$ & $-0.31 \pm 0.06$ \\
2001 PK47 & $-0.15 \pm 0.07$ & $-1.02 \pm 0.10$ & $-0.09 \pm 0.11$ \\
2001 QC298 & $-0.43 \pm 0.06$ & $-1.03 \pm 0.07$ & $-0.26 \pm 0.08$ \\
2001 QR297 & $0.048 \pm 0.06$ & $-0.98 \pm 0.08$ & $-0.16 \pm 0.11$ \\
2002 PD155 & $-0.24 \pm 0.15$ & $-0.71 \pm 0.17$ & $-0.36 \pm 0.16$ \\
2003 LD9 & $0.051 \pm 0.06$ & $-1.08 \pm 0.08$ & $-0.24 \pm 0.09$ \\
2003 QF113 & $0.020 \pm 0.07$ & $-1.14 \pm 0.10$ & $-0.15 \pm 0.12$ \\
2003 SQ317 & $-0.51 \pm 0.06$ & $-1.41 \pm 0.10$ & $-1.37 \pm 0.24$ \\
2003 UZ117 & $-0.47 \pm 0.02$ & $-1.52 \pm 0.03$ & $-1.22 \pm 0.07$ \\
90568 (2004 GV9) & $-0.13 \pm 0.01$ & $-1.06 \pm 0.01$ & $-0.25 \pm 0.01$ \\
145452 (2005 RN43) & $-0.13 \pm 0.01$ & $-1.10 \pm 0.01$ & $-0.24 \pm 0.01$ \\
\multicolumn{4}{c}{Cold Classical}  \\ \hline
1998 WY24 & $-0.07 \pm 0.07$ & $-0.96 \pm 0.10$ & $-0.27 \pm 0.11$ \\
2000 CE105 & $0.091 \pm 0.09$ & $-0.99 \pm 0.11$ & $-0.17 \pm 0.13$ \\
123509 (2000 WK183) & $-0.08 \pm 0.05$ & $-1.03 \pm 0.07$ & $-0.23 \pm 0.08$ \\
2000 WT169 & $-0.07 \pm 0.04$ & $-0.95 \pm 0.05$ & $-0.29 \pm 0.06$ \\
2000 WW12 & \nodata & \nodata & \nodata \\
88268 (2001 KK76) & $0.072 \pm 0.06$ & $-0.88 \pm 0.07$ & $-0.15 \pm 0.07$ \\
160147 (2001 KN76) & $-0.06 \pm 0.06$ & $-1.08 \pm 0.08$ & $-0.23 \pm 0.10$ \\
2001 OQ108 & $0.065 \pm 0.08$ & $-1.00 \pm 0.11$ & $-0.23 \pm 0.13$ \\
2001 QS322 & $-0.01 \pm 0.08$ & $-1.04 \pm 0.10$ & $-0.15 \pm 0.12$ \\
Teharonhiawako (2001 QT297) & $-0.07 \pm 0.05$ & $-1.05 \pm 0.06$ & $-0.19 \pm 0.06$ \\
2001 QX297 & $0.066 \pm 0.06$ & $-1.01 \pm 0.08$ & $-0.29 \pm 0.09$ \\
2001 RW143 & $-0.10 \pm 0.10$ & \nodata & \nodata \\
2001 RZ143 & $-0.07 \pm 0.05$ & $-0.99 \pm 0.06$ & $-0.27 \pm 0.07$ \\
2002 CU154 & $-0.01 \pm 0.08$ & $-1.10 \pm 0.11$ & $-0.41 \pm 0.14$ \\
2002 FW36 & \nodata & \nodata & $-0.31 \pm 0.14$ \\
2002 PV170 & $0.004 \pm 0.04$ & $-1.02 \pm 0.05$ & $-0.26 \pm 0.06$ \\
2003 GH55 & $-0.02 \pm 0.04$ & $-0.88 \pm 0.04$ & $-0.25 \pm 0.05$ \\
2003 HG57 & $-0.04 \pm 0.05$ & $-1.19 \pm 0.07$ & $-0.04 \pm 0.08$ \\
2003 QZ111 & \nodata & \nodata & \nodata \\
2006 HW122 & $-0.05 \pm 0.10$ & $-0.93 \pm 0.13$ & $-0.48 \pm 0.18$ \\
\enddata
\tablenotetext{b}{(F606w-F814w) colour determined from observations presented by \citet{Benecchi2011} (see Section~\ref{sec:observing}).}
\end{deluxetable}

\appendix
\section{Minimum Spanning Tree Clustering and the \FOP}

Various methods exists to identify clusters within multivariate data. Here we consider the use  of minimum spanning tree based clustering methods. A minimum spanning tree (MST) is an acyclic connected graph of a set of data points which minimizes some weight between pairs of data points. Most common MSTs are Euclidean. That is, a Euclidean minimum spanning tree is the graph which minimizes the total Euclidean distance of the graph's edges. In its most general sense, MST based clustering algorithms remove or prune {\it inconsistent} edges from the graph, resulting in subgraphs, or clusters of the data where the removal of $k$ edges results in $k+1$ clusters. The concept of this approach is simple; if a data set is truly clustered, then only one edge will exist in the MST connecting one cluster to the next.

The key to MST based clustering lies in the definition of {\it inconsistent}. While many criterion have been proposed \citep[see][for a review]{Jain1999}, perhaps the most intuitive is the edge-length based criterion proposed by \citep{Zahn1971} which simply rejects the largest edges in a tree. While conceptually simple, this method can suffer from chaining between separate clusters preventing those clusters from being identified, or the artificial identification of a cluster caused by some unusually distant data point. Other alternative methods which prune based on deviations about cluster means are also conceptually simple, but avoid much of the problems faced by largest edge methods. We adopt the method proposed by \citet{Grygorash2006} which prunes those edges that result in the largest decreases in the weighted standard deviations of cluster members about their cluster means. \citet{Grygorash2006} calls this the maximum standard deviation reduction clustering algorithm, or \textbf{MSDR}, and has shown this method to be successful in correctly extracting otherwise complex structures in artificial data.

Two approaches exist to determine the number of clusters in a dataset. The first is simply to set the number of clusters {\it a priori}. Clearly, this requires prior information about the underlying structure of the data and presents an undesirable situation if that information is unavailable. Alternatively, some stopping criterion can be defined. The stopping criterion  chosen by \citet{Grygorash2006} adds additional clusters until the first local minimum in the curve of weighted standard deviation versus $k$ is found, thus providing the {\it most desirable} number of clusters. 

One question still remains with the use of cluster methods that utilize stopping criterion - are the clusters real or artificial? Most multivariate clustering methods do not assign a significance to the existence of identified clusters. Here we present a novel approach to the problem with the creation of a new statistic which can be compared to the an artificial distribution to generate a significance level for that statistic.

Some of the most easily identified clusters are those which are well separated in the dataset - clustering by longest edge, or by average cluster separation have been proposed as a clustering metrics \citep{Zahn1971,Hartigan1981}. Distance alone however, is insufficient. A cluster must additionally be well sampled before its identification may be believed. This led to the RUNT test of multi modality by \citet{Hartigan1992} in which the smallest cluster population size is used as the test statistic, albeit in hierarchical clustering rather than in the MST-based clustering we consider here. It follows that these two metrics, separation and size could be combined. We propose that a good statistic for quantifying the significance of division of a dataset into clusters might be the size of the purported clusters weighted by the distance between them, the latter of which can be found from the pruned edge in the MST. Thus, we define a new statistic as follows. For a set of $N$ points $S$ in $\mathbb{E}^n$ connected by a Euclidean MST $T$, we divide $T$ into two clusters or subtrees $t_1$ and $t_2$  by pruning some edge of $T$ with length $l$. We define the $F$ statistic of the pruned edge as

\begin{equation}
F=2\left(\frac{{\min{\left( |t_1,|t_2| \right)}}}{N}\right) \left(\frac{l}{l_{max}}\right)
\label{eq:F}
\end{equation}

\noindent
where $|t_i|$ is the size of the cluster $t_i$, and $l_{max}$ is the Euclidean magnitude of the longest edge in $T$. $F$ can take on any value in the range $(0,1]$. The closer the value is to 1, the more the sample $S$ appears bifurcated. Clearly, such a statistic gives higher weights or biases to divisions which create the largest cluster sizes, a property of many different clustering techniques \citep{Everitt2011}. While not appropriate in all cases, we feel this metric better reflects reliable divisions by avoiding the creation of many small and unrealistic sub clusters.

As it is written in Equation~\ref{eq:F}, the $F$ statistic can be applied to any MST-based clustering method, regardless of the measure of {\it inconsistency} or the stopping criterion. Alternatively, $F$ can be combined with additional information to be used as a test for the existence of a bifurcation in a dataset. The trick comes from the expectation that an easily identified cluster is one that is well isolated from other clusters, not just at a single point, but across its entire extent. It follows that a surface can be found which optimally divides the regions occupied by two separate clusters. In many cases, the simplest surface is a flat one. That is, a line in $\mathbb{E}^2$, a plane in $\mathbb{E}^3$, etc. Clearly, the flat surface which optimally divides  two clusters must intersect with the MST edge that divides them. In addition, this flat surface, which may not be unique, will have fewest members of each cluster on the wrong side of that plane. 

We define the {\it optimal plane} of two sub clusters $t_1$ and $t_2$ found by pruning edge $e$ from a MST of the dataset $S$, as any flat surface which intersects that edge and has the lowest $N_o=n_{o,1}+n_{o,2}$, where $n_{o,i}$ is the number of members of cluster $t_i$ on the wrong side of the optimal plane. We note that the problem can be further simplified by requiring that the optimal surface contain the midpoint of edge $e$, a simplification we adopt.

The {\it optimal plane} can be used in conjunction with $F$ to find the most {\it inconsistent} edge which proves most optimal in dividing a data set. That is, the edge which, upon pruning, creates the two most well separated clusters is the one which has the {\it optimal plane} with smallest $N_o$ and largest $F$. The procedure is then to probe each edge in the MST to find those with the smallest $N_o$, then choose that edge with maximizes $F$. One distinct disadvantage of the use of the {\it optimal plane} is the difficulty in identifying the full extent of a curved cluster if another cluster is contained within its curvature. Clearly, the use of the {\it optimal plane} is not always appropriate. But its simple interpretation makes its use desirable in many situations. 

The significance of the resultant division can be determined by two separate methods. The first and most preferable method is to bootstrap the dataset by resampling each dimension of the data independently producing an artificial sample of the same size as the original data. The {\it optimal plane} and resultant value of $F$ and $N_o$ are then determined from this artificial dataset, and the process repeated to produce the distribution of $F$ and $N_o$ sampled from the observed distribution of the data. The probability of finding an edge with an $F$-value equal to or larger than that found from the dataset {\it and} and $N_o$-value equal to or smaller than that found from the dataset are determined giving the probability of the null hypothesis. That is, the probability that the dataset is consistent with being drawn from a single population distributed like the observed data.

Alternatively, the observed value of $F$ and corresponding $N_o$ can be compared against the uniform distribution. Formally, this approach tests a different null hypothesis than the bootstrapping method, that the observed $F$ and corresponding $N_o$ are consistent with the distribution drawn from the uniform distribution. This approach however, is more advantageous than bootstrapping for small datasets where bootstrapping can produce a biased result when the underlying distribution is not well sampled. With either calibration approach, the combined usage of $F$ and the {\it optimal plane} provides a test of the existence of a bifurcation in the observed data - we call this the \FOP. We reiterate however, for a sufficient sample size - initial tests suggest a sample of ten or more data points - the bootstrapping approach is nominal.

To show the performance of \textbf{MSDR} clustering technique and the \FOP, we present three test cases. For simplicity, we present all test cases in $\mathbb{E}^3$. The first case is a simulated dataset made up of two equally sampled normal distributions with different means. The second case is reminiscent of the observations we present in this manuscript. Two different populations are drawn from two lines of different slopes. The last is a special case in which the \FOP should have difficulty. One cluster is drawn from a normal distribution while the second is drawn from an annulus centred on and surrounding the normal distribution. In all cases, the sample size is 50, equally split between the two clusters. Two examples at different separations are tested in all cases.

\begin{figure}[h] 
   \centering
      \epsscale{0.8}
   \plotone{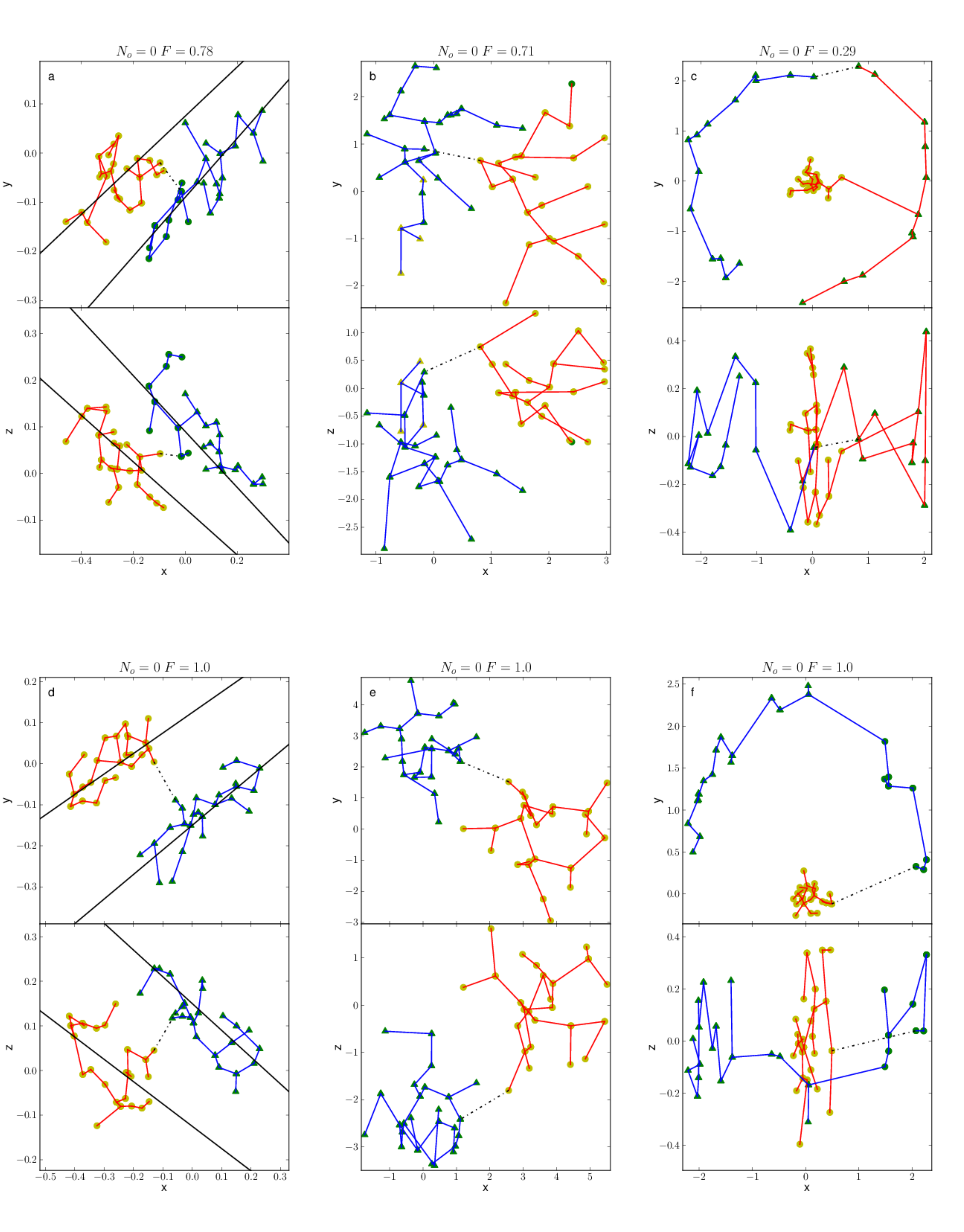} 
   \figcaption{Simulated tests of \textbf{MSDR} clustering and the \FOP. Triangles and circles represent objects randomly drawn from clusters 1 and 2 respectively. The edges of the minimum spanning tree are shown as line segments. Green and yellow points are members of the first and second samples identified by \textbf{MSDR}. The segment pruned by the \FOP - that for which the {\it optimal plane} with maximal $F$ is found - is shown as the dashed line. The red and blue segments show the two sub clusters as a result of the \FOP division and the resultant $N_o$ and $F$ are displayed. \textbf{a,d}: samples drawn from two close and distant lines respectively. The lines from which the points were sampled are shown. \textbf{b,d}: samples drawn from 3D normal distributions with distance between the normal means of 2 and 5 respectively. \textbf{c}: One cluster drawn from a 3D gaussian with standard deviation $\frac{1}{3}$ and the other drawn from a circle with radius 1. \textbf{e}: as in \textbf{c} but for a half circle.
   \label{apfig:frames}}
\end{figure}

The results of the \textbf{MSDR} clustering technique and the \FOP are shown in Figure~\ref{apfig:frames}. The probability of null hypothesis associated with the presented tests were determined from the bootstrapping approach, and are $<2$\% for that shown in \textbf{a,b,c,d,e}  and $40$\% for \textbf{f}.

These tests reveal that, in general, the \FOP performs as well, or even better than the \textbf{MSDR} clustering. That is, where the two clusters are well separated, both techniques produce identical results. When the clusters are closer together however, the \FOP more frequently identifies the correct edge from which to separate the two clusters, resulting in fewer mis-classifications of cluster membership. The second case, that of two lines, is particularly instructive. In the situation where the two samples were generated close together (Figure ~\ref{apfig:frames}\textbf{a}), the \FOP typically outperformed \textbf{MSDR} in the sense that the \FOP correctly determined cluster membership for most or all data points much more often than did \textbf{MSDR}, even in cases with low values of $F$.

As expected however, the \FOP performs poorly in the third case in which one cluster wraps about the other. A typically result was a division of the arc into two sub clusters with very low $F$, despite the fact that the correct cluster distribution is clear, and easily identified by the \textbf{MSDR} clustering. In the case of a half arc, the \FOP could correctly identify the two clusters with very high $F$, though with low frequency. This result demonstrates the primary weakness of the \FOP in that the test implicitly assumes the clusters can be well separated by some plane.

The performance of the \FOP was also significantly diminished when the two generated clusters were not of the same sample size. The tests shown in Figure~\ref{apfig:frames}\textbf{a} and \textbf{d} were performed with 35 and 15 data points in each generated cluster. For the case \textbf{a} where the lines are furthest separated, the \FOP performed well, correctly identifying the bifurcation, albeit with higher probability of the null hypothesis. For case \textbf{d} with a smaller separation between the lines, the \FOP failed the majority of time. This result demonstrates the second weakness of the \FOP which weights the results to clusters of nearly equal population. In situations where both weaknesses are avoided however, the \FOP performs quite well, even outperforming the successful \textbf{MSDR} clustering technique.

In addition it should be noted that the \FOP was not driven by some mathematical properties of minimum spanning trees. Rather, the $F$ statistic is an ad hoc combination of two previously suggested clustering metrics, the properties of which remain mathematically unproven. The use of the \FOP bears this additional caveat. Despite this however, our simulations demonstrate the ability of the \FOP to identify the correct clusters at least as well as other techniques.

\end{document}